\documentclass[fleqn,usenatbib]{mnras}
\usepackage{newtxtext,newtxmath}
\usepackage[T1]{fontenc}
\usepackage{ae,aecompl}
\usepackage{graphicx}	% Including figure files
\usepackage{amsmath}	% Advanced maths commands
\usepackage{amssymb}	% Extra maths symbols
\usepackage{gensymb}

\usepackage{pslatex}
\title[Variability of HD~27463]{Rotational and pulsational variability in the \textit{TESS} light curve of HD~27463}
\author[Khalack et al.]
       {
        \parbox{\linewidth}{\flushleft V. Khalack$^1$, C. Lovekin$^2$, D. M. Bowman$^3$, O. Kobzar$^1$, A. David-Uraz$^4$, E. Paunzen$^5$, J. Sikora$^{6, 7}$, P. Lenz$^8$, O. Kochukhov$^9$, D. L.  Holdsworth$^{10}$, G. A. Wade$^{7}$ }  \\
        \\
          $^1$D\'epartement de Physique et d'Astronomie, Universit\'e de Moncton, Moncton, N.B., Canada E1A 3E9\\
          $^2$Department of Physics, Mount Allison University, Sackville, N.B., Canada E4L 1E6\\
          $^3$Institute of Astronomy, KU Leuven, Celestijnenlaan 200D, 3001 Leuven, Belgium\\
          $^4$Department of Physics \& Astronomy, University of Delaware, 217 Sharp Lab, Newark, DE 19716, USA\\
          $^5$Department of Theoretical Physics and Astrophysics, Masaryk University, Kotl\'a\v{r}sk\'a 2, 611\,37 Brno, Czech Republic\\
          $^6$Department of Physics, Engineering Physics \& Astronomy, Queen's University, Kingston, ON, Canada K7L 3N6\\
          $^7$Department of Physics and Space Physics, Royal Military College of Canada, PO Box 17000 Kingston, ON, Canada K7K 7B4\\
          $^8$Ronin Institute, Montclair, NJ 07043, USA\\
          $^9$Department of Physics and Astronomy, Uppsala University, Box 516, 75120, Uppsala, Sweden\\
          $^{10}$Jeremiah Horrocks Institute, University of Central Lancashire, Preston PR1 2HE, UK\\
            }
\date{Accepted ???.
      Received ???;
      in original form ???}

\pagerange{\pageref{firstpage}--\pageref{lastpage}} \pubyear{2019}

\begin{document}

\maketitle

\label{firstpage}

\begin{abstract}
The new photometric data on pulsating Ap star
HD~27463 obtained recently with the Transiting Exoplanet Survey Satellite (\textit{TESS}) are analysed to search for variability.
Our analysis shows that HD~27463 exhibits two types of photometric variability. The low frequency variability with the period $P$ =~2.834274 $\pm$ 0.000008 d can be explained in terms of axial stellar rotation assuming the oblique magnetic rotator model and presence of surface abundance/brightness spots,
while the detected high-frequency variations are characteristics of $\delta$~Scuti pulsations.
From the analysis of Balmer line profiles visible in two FEROS spectra of HD~27463 we have derived its effective temperature and surface gravity, finding values that are close to those published for this star in the \textit{TESS} Input Catalogue (TIC). Knowing the rotation period and the v$\sin{i}$ value estimated from the fitting of Balmer line profiles we found that the rotational axis is inclined to the line of sight with an angle of $i=33\pm8\deg$.
Our best-fitting model of the observed pulsation modes results in an overshoot parameter value $f_{ov} = 0.014$ and values of global stellar parameters that are in good agreement
with the data reported in the TIC and with the data derived from fitting Balmer line profiles. This model indicates an age of 5.0 $\pm$~0.4 $\times 10^8$~yrs, which corresponds to a core hydrogen fraction of 0.33.
\end{abstract}


\section{Introduction}

Ever since a strong peculiarity was discovered in the Si\,{\sc ii} doublet of the archetypal star $\alpha^2$ CVn \citep{Maury+Pickering97}, chemically peculiar stars have been the subject of numerous studies. In particular, Ap stars (classified as `CP2' in contrast to Am, HgMn or He stars) are known to exhibit periodic brightness variations generally understood to be associated with rotation taking into account that in these stars, strong magnetic fields \citep{Babcock58} lead to the creation of surface abundance patches.

A subset of Ap stars lies within the $\delta$ Scuti instability strip. Accordingly,  high-frequency pulsations have been discovered in a number of Ap stars, starting with Przybylski's star (HD 101065; \citealt{Kurtz78}). Dedicated surveys have revealed a few rapidly oscillating Ap (or `roAp') stars, but were limited to  relatively %fairly
high-amplitude pulsations since they were carried out using ground-based observations.  However, ground-based surveys and the advent of high-precision spaced-based photometry have enabled the discovery and detailed study of new Ap pulsators (e.g. HD~24355; \citealt{Holdsworth+14,Holdsworth+16}) and the revisiting of known pulsators (e.g. HD~137949; \citealt{Kurtz82,Holdsworth+18}).
%However, the ground-based surveys %(e.g. \citealt{Holdsworth+14}),
%and the advent of high-precision space-based photometry {\bf have} enabled the discovery and detailed study of new Ap pulsators (e.g. HD~24355 and HD~137949; % discovered using K2 observations;
%\citealt{Kurtz82,Holdsworth+14,Holdsworth+16,Holdsworth+18}).

The Transiting Exoplanet Survey Satellite (\textit{TESS}) was launched by NASA on  2018 April 18 with the purpose of detecting exoplanets using the transit method \citep{Ricker+15}. Over the course of its  2-yr nominal mission, it will cover most of the sky in 26 sectors with its four wide-field cameras, each covering $24 \times 24\deg$. Each sector is observed for $\sim$27~d, and sectors in each ecliptic %celestial
hemisphere overlap near the poles. Therefore stars observed by \textit{TESS} will have temporal baselines between 27~d and about a year. All fields are observed with a 30-minute cadence (in full-frame images, or FFIs) but a subset ($> 200,000$) of stars  has been selected for short-cadence (2 min) observations, forming the candidate target list (CTL; \citealt{Stassun+18,Stassun+19}). With such a temporal baseline and cadence, \textit{TESS} not only represents a formidable asset for exoplanetary detection, but can also be leveraged very profitably for asteroseismology (e.g. \citealt{Campante+16}). As such, this mission represents a tremendous opportunity to not only significantly increase the sample size of known pulsating Ap stars (see e.g. \citealt{Cunha+19}), but also, given the exquisite data quality, to perform a detailed characterization of specific objects of interest.

The chemically peculiar (CP) star HD~27463 (TT~Ret, HIP~19917, Renson~7050, TIC~38586082) is known as an $\alpha^2$ CVn type variable with spectral type  Ap\,EuCr(Sr) \citep{Houk+Cowley+75}. Photometric variability of this star with a period of $P = 2.835$~days %2.83507 $\pm$ 0.00008 days and the epoch of %zero phase
%E = 2448502.2944$\pm$0.0004
has been reported in the Hipparcos and Tycho catalogues \citep{ESA97} and confirmed by \citet{Renson+Catalano+01}.
%{\bf Similar period of photometric variability P = 2.835 days has been derived for this star by
%Hipparcos catalogue (the New Reduction)} \citep{vanLeeuwen07}.

HD~27463 is a long period ($\sim$370 yr) visual binary  with a separation of $\sim$0.3 arcsec and a magnitude difference between the Ap primary and the secondary of about 0.43 in the $V$ band (e.g. \citealt{Baize+Petit89} and references therein) and 0.9 in the Str\"{o}mgren $y$ band \citep{Hartkopf+12}. Taking into account the observed small angular separation between the two components, they both contribute to the flux detected by \textit{TESS}.

Previous surveys designed to detect high-frequency pulsations did not yield any detection in this star \citep{Martinez+Kurtz94, Joshi+16}. However, using \textit{TESS} data, \citet{Cunha+19} and \citet{Sikora+19} have classified HD~27463 as a suspected new $\delta$~Scuti variable. Therefore, in this study, we set out to perform a more involved investigation of this star's \textit{TESS} photometry, in conjunction with high-resolution spectroscopy. This study is carried out as part of the MOBSTER project, an analysis of magnetic O, B and A stars  using \textit{TESS} light curves \citep{David-Uraz+19}, and of the VeSElkA project, a search for CP stars with vertical stratification of  elemental abundance \citep{Khalack+LeBlanc15a,Khalack+LeBlanc15b}.

The photometric and spectral observations and the respective reduction procedures %used to reduce the \textit{TESS} data
are described in Section~\ref{obs}. The use of automatic software to analyse the light curves is considered in Section~\ref{lc}. Comparison of the fundamental stellar parameters derived from the analysis of Balmer line profiles and from the simulation of stellar pulsations are shown in Section~\ref{parameters}. % and Section~\ref{Balmer_par} respectively.
The
%results on photometric variability are presented in Section~\ref{analysis}, while a
discussion follows in Section~\ref{discus}.

%Observation of CP stars with TESS.

\section{Observations and data reduction}
\label{obs}

%\subsection{Reduction of TESS light curves}
%\label{lc}

\begin{figure}
\begin{center}
\includegraphics[width=1.2in,angle=-90]{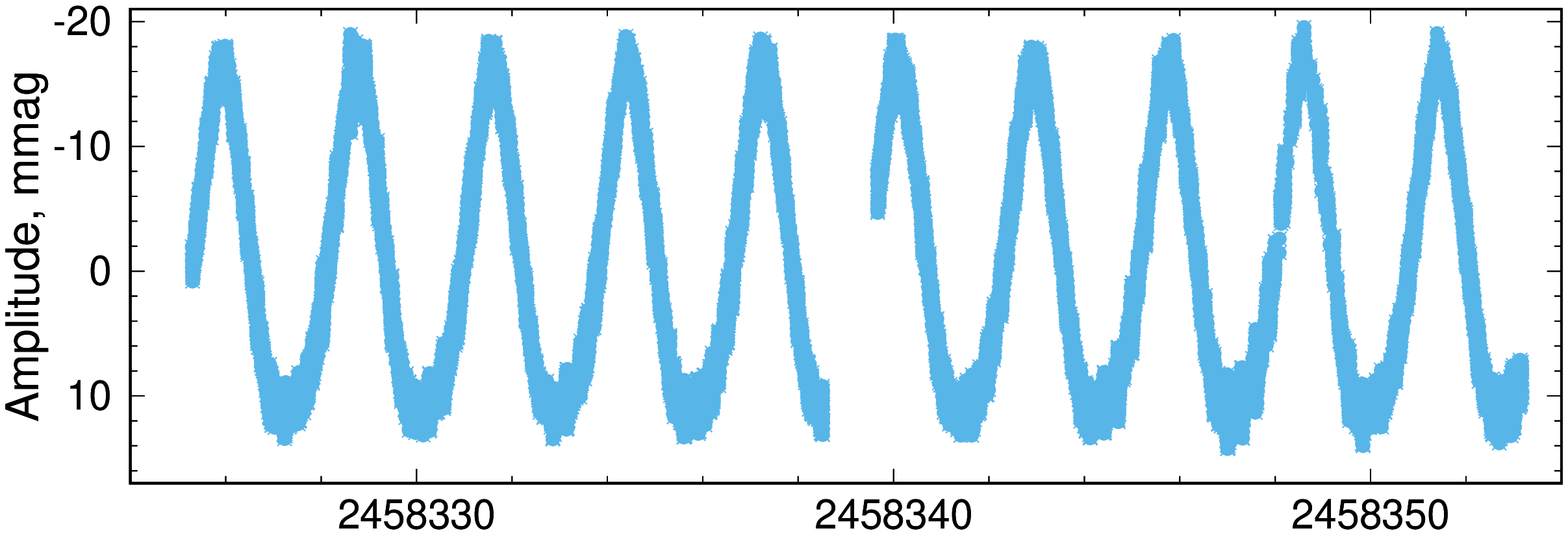} \\
\includegraphics[width=1.2in,angle=-90]{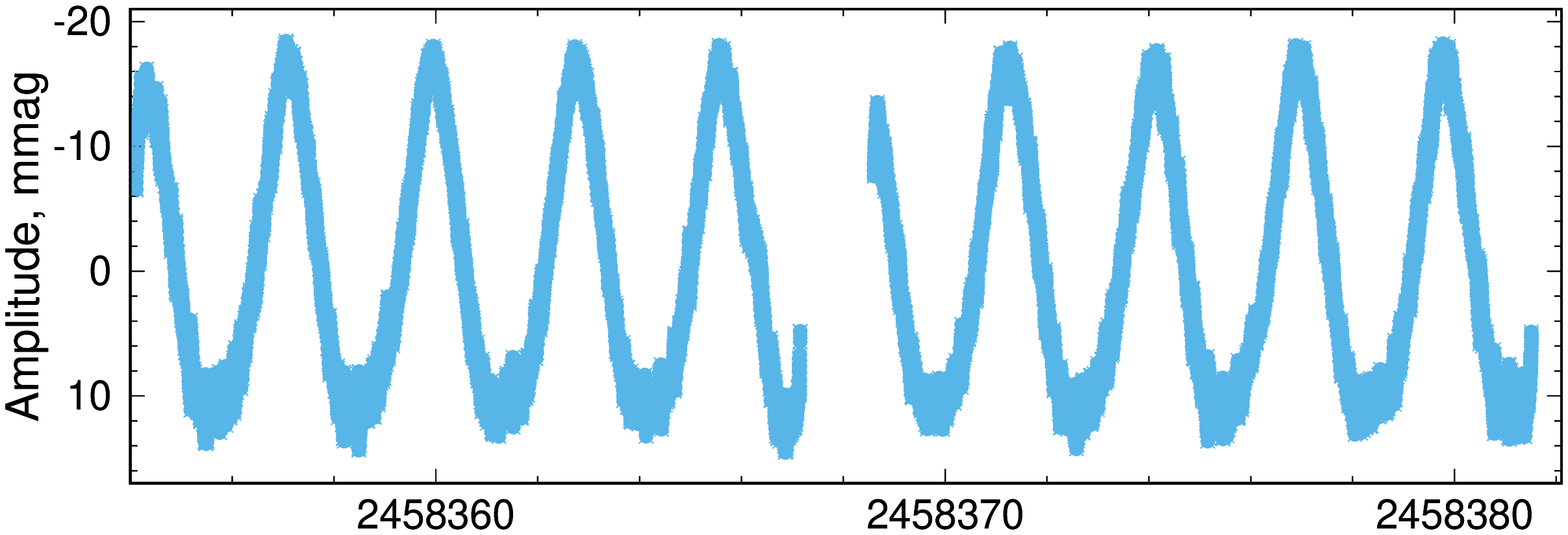} \\
\includegraphics[width=1.2in,angle=-90]{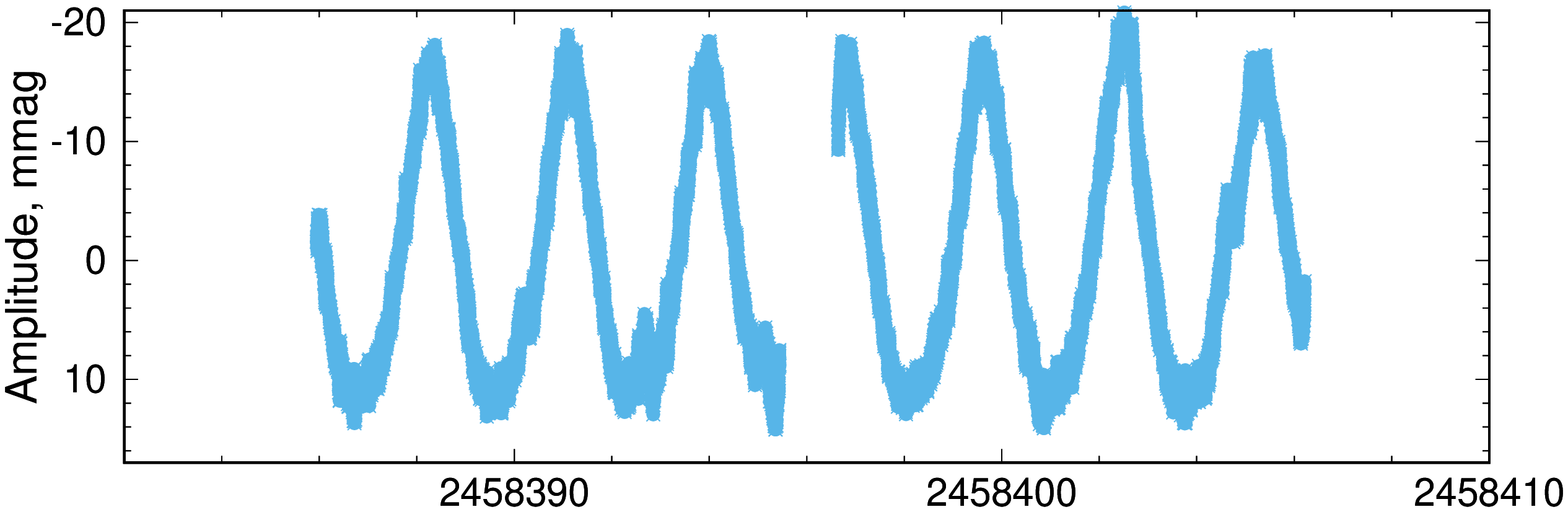} \\
\includegraphics[width=1.2in,angle=-90]{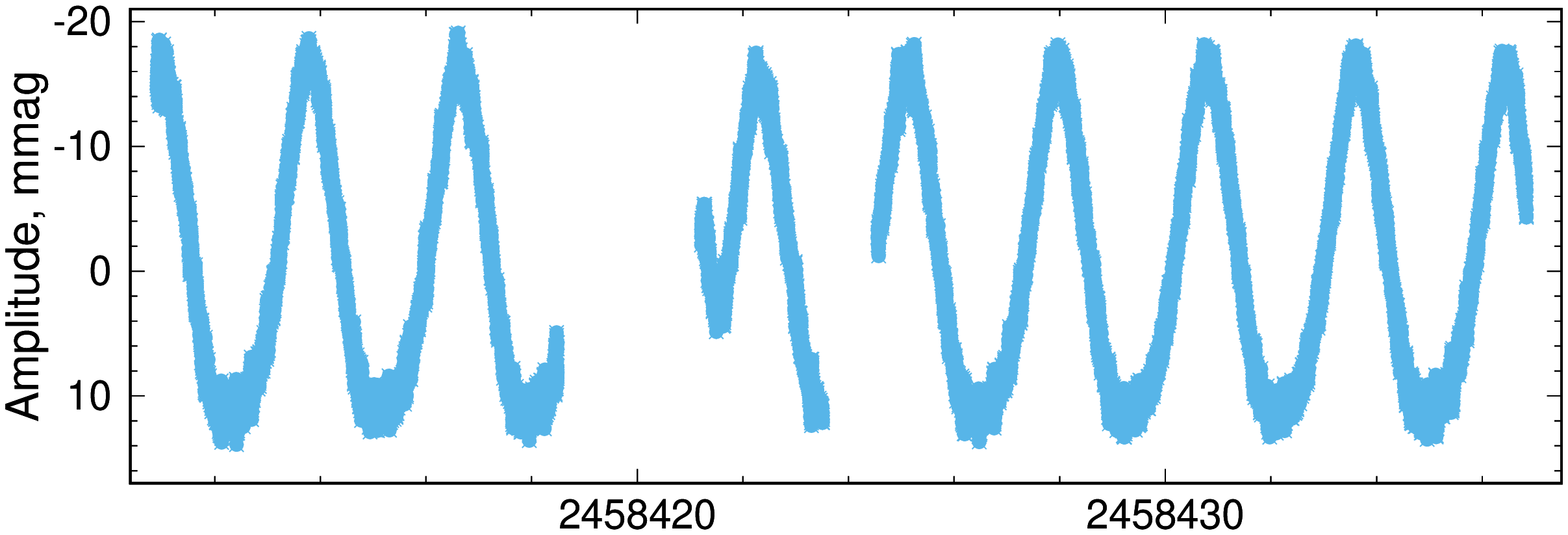} \\
\includegraphics[width=1.3in,angle=-90]{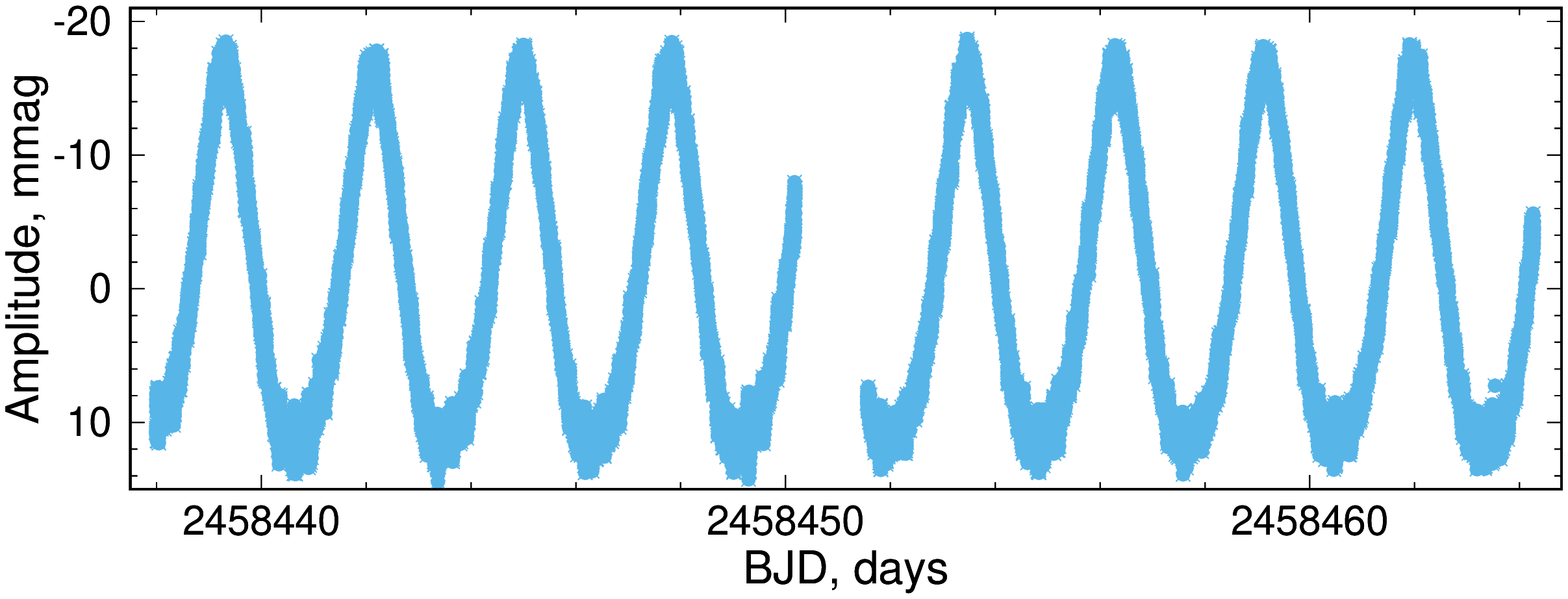} \\
\caption{Examples of the detrended light curves obtained by \textit{TESS} for HD~27463 in Sectors 1-5 (from top to bottom). }
\label{fig1a}
\end{center}
\end{figure}

\begin{figure}
\begin{center}
\includegraphics[width=1.3in,angle=-90]{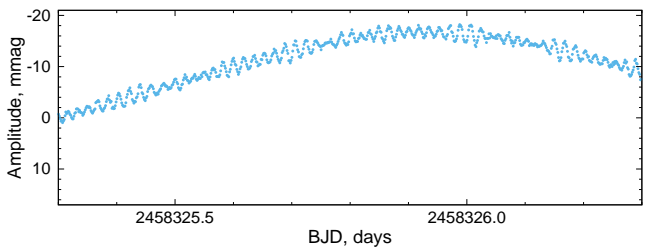} \\
\caption{Part of the light curve obtained with \textit{TESS} for HD~27463 over a time span of 1 day in Sector 1. The observed long-term variability is related to the stellar rotation (see Fig.~\ref{fig1a}). Stellar pulsations with frequencies expected from a $\delta$ Scuti variable and beating of close pulsation frequencies are clearly visible (see also the lower panel in Fig.~\ref{fig4}). }
\label{fig1b}
\end{center}
\end{figure}

HD~27463 has been observed by \textit{TESS} in Sectors 1--5 in the frame of the \textit{TESS} guest program \textit{Rotationally-Induced Variability Of Chemically Peculiar (CP) Stars} \citep{Ricker+Vanderspek}. The \textit{TESS} data products have been reduced employing the \textsc{spoc} pipeline \citep{Jenkins+16}, which is based on the \textit{Kepler} Science Processing Pipeline.
The light curves and the associated data listed in the \textit{TESS} Input Catalogue (TIC\footnote{https://archive.stsci.edu/tess/tic\_ctl.html}) have been downloaded via the Mikulski Archive for Space Telescopes (MAST\footnote{http://archive.stsci.edu/tess/all products.html}) and are publicly available. The extracted time series of flux measurements taken at different Barycentric Julian Dates (BJD) were transformed to units of stellar magnitude. The light curves have been normalised to a mean of zero  (see Figs.~\ref{fig1a},~\ref{fig1b}).
%During the Fourier Transformation (FT) of the light curves the mean value has been subtracted to study their variability around zero and to perform an analysis of present periodic signals.

To accurately determine the rotation period and the pulsation mode frequencies of HD~27463, we further detrended the 2-min \textit{TESS} light curve following the methodology of \citet{Bowman+18b}. We fitted a periodic rotational modulation model comprising multiple harmonics of the rotation frequency, $\nu_{\rm rot} = 0.352824 \pm 0.000001$~d$^{-1}$, to the light curve using non-linear least-squares. We subsequently subtracted this preliminary fit and modelled the residuals using a locally weighted scatterplot smoothing (LoWeSS) filter \citep{Cleveland79,Seabold+Perktold10}. To create our final detrended light curve, we subtracted the modelled residuals from the original light curve. This ensures that a high-quality light curve is used for the determination of a rotation period and for the extraction of pulsation mode frequencies, since the long-term trends and instrumental systematics have been removed.

%Add description of the FEROS spectra.
HD~27463 was observed twice in 2008 with the FEROS spectrograph installed on the 2.2-m Max Planck Gesellschaft/European Southern Observatory (MPG/ESO) telescope in La Silla. The FEROS spectrograph provides high-resolution spectra with R$\sim$ 48000 in the spectral region from 3600 to 9200\,\AA\, \citep{Kaufer+99}. Two available spectra of HD~27463 were reduced with the ESO automatic reduction pipeline employing the barycentric velocity correction. %and the correction of the telluric lines.
%Nowadays
 The spectra are public and have been downloaded from the ESO archive\footnote{http://archive.eso.org/wdb/wdb/adp/phase3\_spectral/query}.
To carry out analysis of Balmer line profiles we have used non-normalized spectra (see Section~\ref{parameters}) and therefore no additional reduction procedure has been applied to the downloaded data.

%\section{Analysis procedure}
%\label{analysis}

\section{Photometric analysis}
\label{lc}

To carry out the analysis of \textit{TESS} light curve we have developed an automatic software (several scripts and codes including the code {\sc period04}) that treats time series, performs periodic analysis, and extracts  additional data for a studied star from the public astronomical databases (TIC, SIMBAD, GALAH).  The software creates images and a list of  identified periodicities, and stores all the information in a final report.

%amplitudes axseed residuals more than three times, see Table~\ref{tab2}).

The code {\sc period04} (version 1.2.9) developed by \citet{Lenz+Breger05} has been specially designed to carry out a statistical analysis of large astronomical data sets with significant gaps and is a powerful tool for the identification and evaluation of periodic signals in long time series of stellar flux measurements. It has been recently modified by one of us (PL) to work  in the {\it batch} mode such that it can be launched automatically from a script with a list of specified parameters. The significant peaks in the \textit{TESS} light curve of HD~27463 have been extracted using the standard procedure of iterative pre-whitening and optimised using a multi-frequency least-squares fit procedure \citep{Lenz+Breger05}. This analysis yielded the detection of multiple frequencies with amplitudes having a significant signal-to-noise ratio. Significant peaks are defined as those with an amplitude larger than four times the level of noise remaining %left
after pre-whitening the signal at %the area of
each studied frequency.
%three times the residuals in frequency spectrum.
%Tracking of a new periodic signal has been carried out until its amplitude was approximately times higher than the remaining residuals.

\begin{figure}
\begin{center}
\includegraphics[width=1.2in,angle=-90]{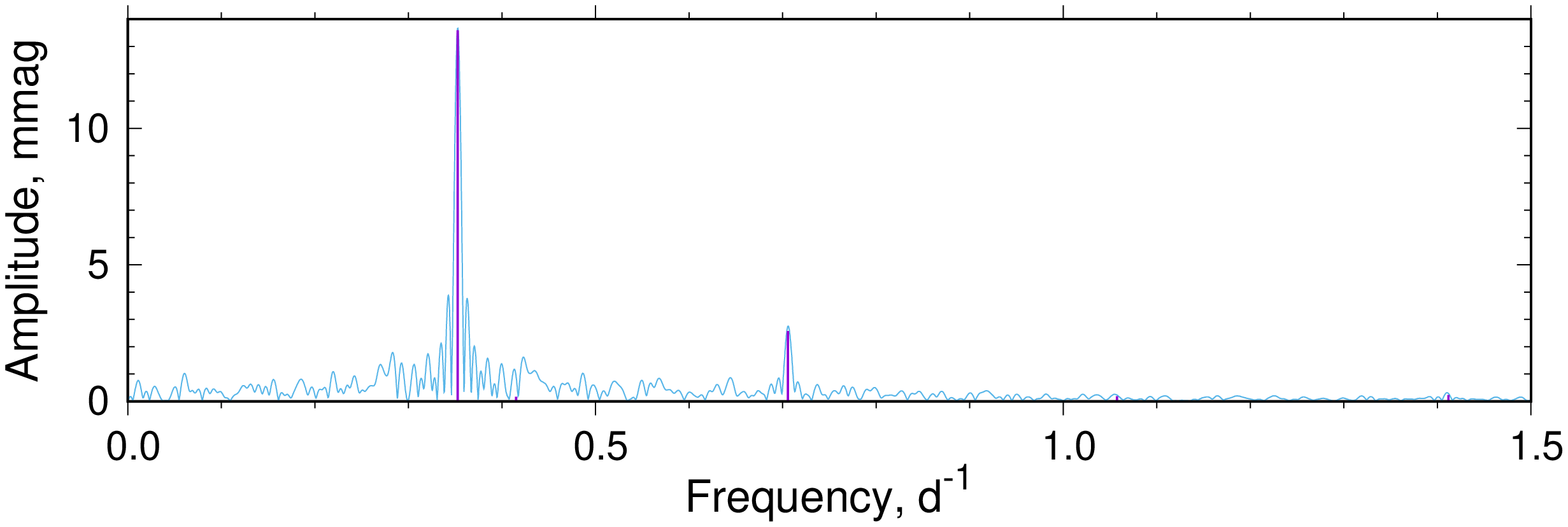} \\
\includegraphics[width=1.2in,angle=-90]{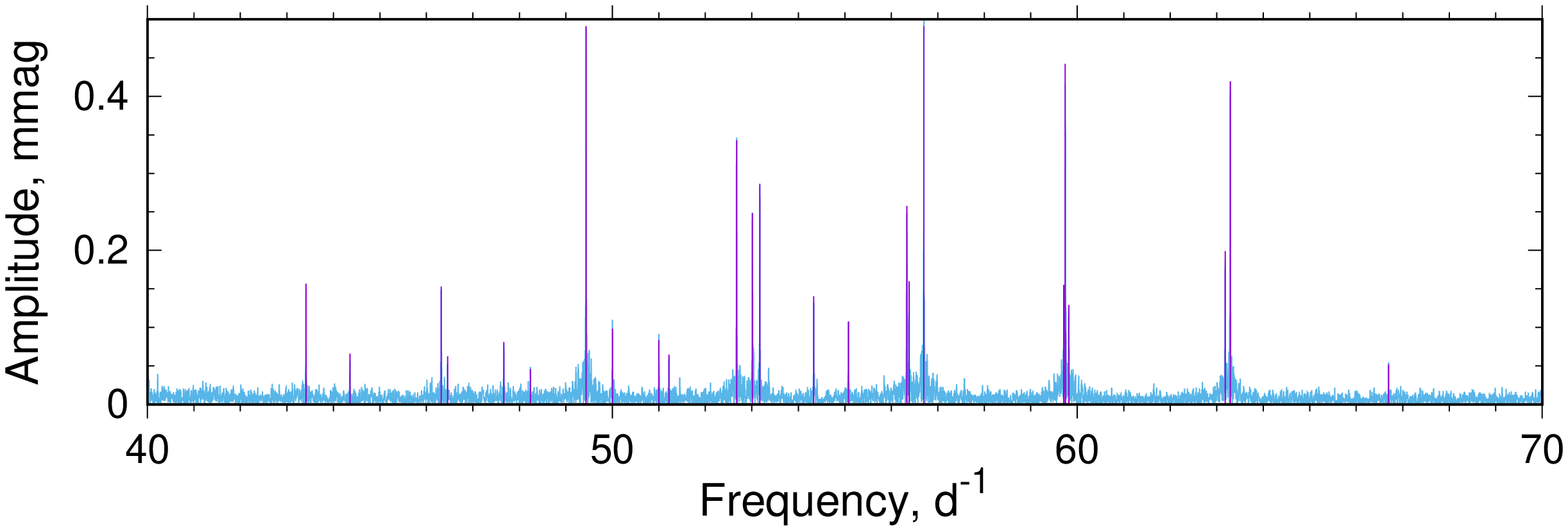} \\
\caption{Examples of amplitude spectra derived from the analysis of HD~27463's light  curve at the low (upper panel) and high (bottom panel) frequencies.
Variability at the low frequencies is related to rotation of the star, while signals detected at the high frequencies are due to stellar pulsations.
 Vertical lines mark the derived frequencies and their amplitudes for the detected periodic signals (see Table~\ref{tab2_All}). }
\label{fig4}
\end{center}
\end{figure}

\begin{table}
\begin{center}
\caption{Characteristics of the periodic signals detected in the light curve of HD~27463 combined from the photometric measurements in the \textit{TESS} sectors 1-5. }
\label{tab2_All}
\def\arraystretch{0.98}
\setlength{\tabcolsep}{3pt}
\begin{tabular}{rccrc}\hline
 Frequency & Amplitude & Phase & S\Big/N & Modes \\ %\multicolumn{3}{c}{Modes}\\
   d$^{-1}$ & mmag &  rad/2$\pi$ &  & $n$ \\ %& $\ell$ & $m$ \\
\hline
$\nu_{rot}$ 0.352824$\pm$0.000001 &  13.596$\pm$ 0.006 & 0.9359$\pm$0.0001 & 29.1 &  \\ %& & \\
% 0.415166$\pm$0.000088 &   0.167$\pm$ 0.006 & 0.6930$\pm$0.0060 & 0.36 &  & \\
2$\nu_{rot}$ 0.705639$\pm$0.000006 &   2.573$\pm$ 0.007 & 0.0895$\pm$0.0004 & 6.6 &  \\ %& & \\
%3$\nu_1$ 1.057342$\pm$0.000078 &   0.188$\pm$ 0.006 & 0.9612$\pm$0.0058 & 0.6 &  & \\
% 1.411708$\pm$0.000062 &   0.238$\pm$ 0.007 & 0.9652$\pm$0.0045 & 1.38 &  & \\
%40.221973$\pm$0.000347 &   0.042$\pm$ 0.006 & 0.3286$\pm$0.0240 &  3.8 &  & \\
%43.409828$\pm$0.000094 &   0.157$\pm$ 0.007 & 0.9349$\pm$0.0279 & 14.5 & {\bf 3} & {\bf -2} \\
43.409828$\pm$0.000094 &   0.157$\pm$ 0.007 & 0.9349$\pm$0.0279 & 14.5 & 12\\ %& {\bf 0} & {\bf 0} \\
44.356363$\pm$0.000223 &   0.066$\pm$ 0.013 & 0.9012$\pm$0.0557 &  6.3 & \\ %& {\bf 1} & {\bf -1} \\
%                       &                    &                & {\bf or} & 3 & 0 \\
46.318147$\pm$0.000096 &   0.153$\pm$ 0.007 & 0.8000$\pm$0.0065 & 14.2 & \\ %&  2 & {\bf 0} \\
46.456655$\pm$0.000235 &   0.062$\pm$ 0.006 & 0.6032$\pm$0.0175 &  5.8 & \\ %&  2 & 0 \\
47.667615$\pm$0.000181 &   0.081$\pm$ 0.006 & 0.7426$\pm$0.0127 &  7.6 & \\ %& {\bf 1} & {\bf -1} \\
%                       &                    &                & {\bf or} & 3 & 0 \\
48.238558$\pm$0.000322 &   0.046$\pm$ 0.007 & 0.4482$\pm$0.0228 &  4.1 & \\ %&  1 & 0 \\
%                       &                    &                & {\bf or} & {\bf 3} & {\bf 1} \\
49.433689$\pm$0.000030 &   0.491$\pm$ 0.008 & 0.3511$\pm$0.0022 & 29.4 & \\ %&  2 & -1 \\
50.002473$\pm$0.000149 &   0.098$\pm$ 0.006 & 0.8432$\pm$0.0115 &  5.9 & 14 \\ %&  {\bf 0} & {\bf 0} \\
50.999734$\pm$0.000177 &   0.083$\pm$ 0.007 & 0.7933$\pm$0.0123 &  8.1 & \\ %&  3 & 0 \\
51.215952$\pm$0.000226 &   0.065$\pm$ 0.006 & 0.6727$\pm$0.0147 &  6.6 & \\ %&  {\bf 1} & {\bf -1} \\
52.672629$\pm$0.000043 &   0.343$\pm$ 0.006 & 0.1683$\pm$0.0032 & 18.9 & \\ %&  2 & {\bf -1} \\
53.012605$\pm$0.000059 &   0.248$\pm$ 0.006 & 0.4302$\pm$0.0048 & 14.0 & \\ %&  2 & 0 \\
%53.040666$\pm$0.000329 &   0.045$\pm$ 0.007 & 0.8814$\pm$0.0247 &  2.5 &  & \\
53.173059$\pm$0.000051 &   0.286$\pm$ 0.006 & 0.0920$\pm$0.0038 & 16.0 & \\ %&  2 & 0 \\ %F18
%53.361934$\pm$0.000297 &   0.049$\pm$ 0.006 & 0.6115$\pm$0.0219 &  2.7 &  & \\
54.333293$\pm$0.000105 &   0.140$\pm$ 0.007 & 0.4882$\pm$0.0067 & 13.7 & \\ %&  3 & 0 \\
55.076921$\pm$0.000137 &   0.107$\pm$ 0.007 & 0.7837$\pm$0.0089 &  9.7 & \\ %&  1 & 0 \\
%                       &                    &                & {\bf or} & {\bf 3} & {\bf 1} \\
56.332132$\pm$0.000057 &   0.258$\pm$ 0.006 & 0.8862$\pm$0.0046 & 12.2 & \\ %&  2 & 0 \\
56.337529$\pm$0.000080 &   0.183$\pm$ 0.007 & 0.6321$\pm$0.0054 &  8.6 & \\ %&  2 & 0 \\
56.383938$\pm$0.000092 &   0.160$\pm$ 0.007 & 0.6144$\pm$0.0060 &  7.6 & \\ %&  2 & 0 \\
%56.700889$\pm$0.000030 &   0.491$\pm$ 0.007 & 0.9475$\pm$0.0021 & 23.3 & {\bf 3} & {\bf -2} \\ %F7
56.700889$\pm$0.000030 &   0.491$\pm$ 0.007 & 0.9475$\pm$0.0021 & 23.3 & 16\\ %& {\bf 0} & {\bf 0} \\ %F7
59.709581$\pm$0.000095 &   0.155$\pm$ 0.006 & 0.4958$\pm$0.0066 &  9.4 & \\ %&  2 & {\bf 0} \\
59.741600$\pm$0.000033 &   0.442$\pm$ 0.012 & 0.2427$\pm$0.0043 & 26.7 & \\ %&  2 & 0 \\
59.746637$\pm$0.000155 &   0.095$\pm$ 0.013 & 0.4808$\pm$0.0097 &  5.7 & \\ %&  2 & 0 \\
59.815711$\pm$0.000114 &   0.129$\pm$ 0.007 & 0.7050$\pm$0.0076 &  7.8 & \\ %&  2 & {\bf 0} \\
%63.180209$\pm$0.000074 &   0.199$\pm$ 0.009 & 0.5228$\pm$0.0058 & 12.8 & {\bf 3} & {\bf -2} \\ %F15
63.180209$\pm$0.000074 &   0.199$\pm$ 0.009 & 0.5228$\pm$0.0058 & 12.8 & 18\\ %& {\bf 0} & {\bf 0} \\ %F15
63.290656$\pm$0.000035 &   0.420$\pm$ 0.007 & 0.7557$\pm$0.0025 & 26.9 & \\ %&  2 & {\bf 1} \\
66.696167$\pm$0.000283 &   0.052$\pm$ 0.007 & 0.9038$\pm$0.0236 &  6.5 & \\ %&  {\bf 3} & {\bf -2} \\
\hline
\end{tabular}
\end{center}
\end{table}

Finally, the combined light curve was fitted taking into account all detected frequencies using the least-squares procedure and the obtained frequencies, amplitudes and phases are shown respectively in the first, second and third columns  of Table~\ref{tab2_All}. The presented uncertainties are the largest ones obtained from the error matrix produced by the Levenberg-Marquardt non-linear least-squares fitting procedure and from  a Monte Carlo simulation of a signal with the detected frequencies \citep{Lenz+Breger05}. They do not take into account the estimation errors reported for the flux measurements\footnote{Consideration of the observation errors during the Fourier analysis reveals no significant changes in the derived estimates of frequency or amplitude of the detected signals. Meanwhile it leads to changes of the best fitted phase derived for the weak signals and increases by an order of magnitude the errors obtained from the Monte Carlo simulation for the estimates of frequency, amplitude and phase.}. % \citep{Bevington+69,Lenz+Breger05}.
%In this approach
Using this approach, the uncorrelated uncertainties for the frequency and the phase have been derived \citep{Montgomery+Odonoghue99}.
The 4th column of Table~\ref{tab2_All} contains the signal-to-noise ratio calculated for each mode at the derived frequency based on the simulation of white noise using the dispersion of residual data left after the pre-whitening procedure \citep{Lenz+Breger05}. %Only modes with a signal-to-noise higher than four have been taken into account.
Meanwhile the 5th column of Table~\ref{tab2_All} specifies the radial order of $ell = 0$.
%Note the regularity in the frequency spectrum, i.e. clusters of modes with a general spacing of 3 c/d!
%The significant periodic signals were detected at lower frequencies (see Table~\ref{tab2_All}) and most probably

\begin{figure}
\begin{center}
\includegraphics[width=4.5in,angle=-90]{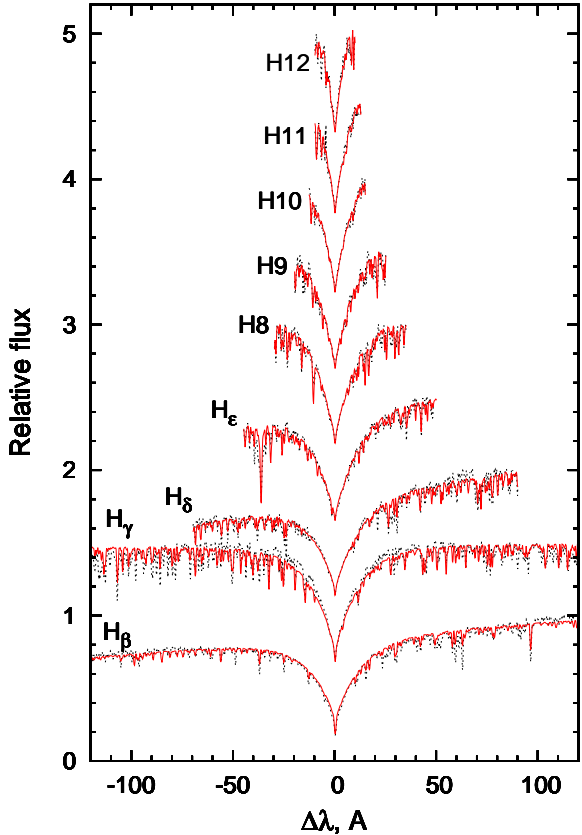} \\
\caption{ An example of fitting the observed Balmer line profiles (thick red line) of HD~27463 with a synthetic spectrum (thin dotted line) that corresponds to $T_{\rm eff}$ = 8700 $\pm$ 100~K, $\log{g}$ = 3.9 $\pm$ 0.1, [M/H]= 0.3 $\pm$ 0.1 ($\chi^2$ = 1.081). The best fit is obtained for the radial velocity v$_{\rm r}$ = 22 $\pm$ 2 km s$^{-1}$ and v$\sin{i}$ = 27 $\pm$ 2 km s$^{-1}$ (see Table~\ref{tab1}).
%Differences between the observed and synthetic profiles are shown at the bottom of this image (in blue).
The observed Balmer line profiles are shifted by 0.5
%and the residuals are shifted by 0.1
for visibility. }
\label{fig3}
\end{center}
\end{figure}

\begin{table}
\begin{center}
\caption{Fundamental stellar parameters derived for HD~27463. }
\label{tab1}
\begin{tabular}{lccc}\hline
 Parameter & TIC,     &  \multicolumn{2}{c}{This article}  \\
           & SIMBAD   & Balmer lines & Pulsations \\
\hline
$T_{\rm eff}$, K      & 8700 $\pm$ 200$^a$ & 8700 $\pm$ 100 & 8800 $\pm$ 100 \\
$\log{g}$             & 3.9 $\pm$ 0.3$^a$  & 3.9 $\pm$ 0.2  & 3.89 $\pm$ 0.05 \\

 [M/H]                &                   & 0.3 $\pm$ 0.1 & \\
v$\sin{i}$, km s$^{-1}$ &                & 27 $\pm$ 2 &   \\
v$_{\rm r}$, km s$^{-1}$ &  26.5 $\pm$ 4.8$^b$ & 22 $\pm$ 2$^c$ & \\
$L_{\rm \star}$, L$_{\rm \odot}$  & 38.7 $\pm$ 10.8$^a$ & & 42.5 $\pm$ 0.6  \\
$R_{\rm \star}$, R$_{\rm \odot}$  &  2.8 $\pm$ 0.4$^a$  & & \\
$M_{\rm \star}$, M$_{\rm \odot}$  &  2.2 $\pm$ 0.4$^a$  & & 2.4 $\pm$ 0.1 \\
v$_{\rm eq}$, km s$^{-1}$ &               & 50 $\pm$ 7  &  47 $\pm$ 5 \\
$i$, $\deg$               &               & 33 $\pm$ 8  & \\
age, $10^8$ yr &         &   & 5.0 $\pm$ 0.4 \\
\hline
\end{tabular}
\end{center}
{\it Notes:} $^a$\citet{Stassun+18,Stassun+19}; $^b$\citet{Gontcharov2006}; $^c$ in barycentric reference frame.
\end{table}

%Our estimates of the effective temperature, surface gravity, metallicity, v $\sin{i}$ and v$_{\rm r}$ have been obtained from the analysis of Balmer line profiles (see Section~\ref{Balmer_par}).

\section{Study of fundamental stellar parameters}
\label{parameters}
Some fundamental stellar parameters of HD~27463 are taken from  public astronomical databases (TIC, SIMBAD\footnote{simbad.u-strasbg.fr/simbad/sim-id?Ident=HIP19917}, see Table~\ref{tab1}).
The effective temperatures reported in TIC has been derived using a new empirical relation between $T_{\rm eff}$ and \textit{Gaia} $G_{\rm BP} - G_{\rm RP}$ colours, based on a set of 19,962 stars \citep{Stassun+19}. The stellar radius was computed using the \textit{Gaia} parallaxes employing the standard expression derived from the Stefan-Boltzmann law. Based on the obtained values of effective temperature and stellar radius one can calculate the respective bolometric luminosity, $L_{\rm bol}$. More details on the procedure of determination of the fundamental stellar parameters reported in TIC and its estimation errors can be found in \citet{Stassun+18,Stassun+19}.

Our estimates of the effective temperature, surface gravity, metallicity, radial velocity and v$\sin{i}$ have been derived by fitting the non-normalised profiles of Balmer lines (see Fig.~\ref{fig3}).
%(except $H_{\rm \alpha}$)
%visible in the two FEROS spectra of this star.
The derived value of v$\sin{i}$ = 27 $\pm$ 2 km~s$^{-1}$ has been confirmed by %preliminary
analysis of several Fe\,{\sc ii} line profiles visible in the two FEROS spectra of this star.
%obtained with the spectrograph FEROS operated at the European Southern Observatory (ESO) in La Silla. %The spectra are public and have been downloaded from the ESO archive\footnote{http://archive.eso.org/wdb/wdb/adp/phase3\_spectral/query}.

We assume here that the high-amplitude periodic variability in the light curve with the lowest frequency and the associated first harmonic (see Table~\ref{tab2_All}) correspond to the stellar %axial
rotation. Therefore, we have used the estimate of the stellar radius from the TIC and the spectroscopic value of v$\sin{i}$ to derive the equatorial velocity, v$_{\rm eq}$, and inclination angle, $i$, of the rotational axis with respect to the line of sight (see Table~\ref{tab1}).

\subsection{Analysis of Balmer line profiles}
\label{Balmer_par}

The Balmer line profiles of HD~27463 have been fitted employing the {\sc fitsb2} code \citep{Napiwotzki+04} to derive best fit parameters ($T_{\rm eff}$, $\log{g}$ and metallicity) of the stellar atmosphere. This code uses grids of synthetic fluxes simulated with the code {\sc phoenix} \citep{Hauschildt+97} for different values of $T_{\rm eff}$, $\log{g}$ and metallicity.
%To derive best fit parameters ($T_{\rm eff}$, $\log{g}$ and metallicity) of stellar atmosphere we have employed the code  \textsc{fitsb2} to perform fitting of the Balmer line profiles.
During this procedure the code takes into account the level of nearby continuum and the intensity of some strong metallic lines that are present in the Balmer wings.
The \textsc{fitsb2} code does not perform an abundance analysis for each chemical element, but it rather results in the estimate of metallicity which serves as an average measure of abundance of metals. For this reason, the spectral lines of most metals are not well fitted in Fig.~\ref{fig3}. %(see residual curves presented at the bottom of this figure).
%for the differences between the synthetic and observed spectra presented at the bottom of this figure).

Specifically, in this study we use the grids
%of synthetic fluxes calculated by \citet{Khalack+LeBlanc15a} with spectral resolution R=60000 for models with different metalicities using the version 15 of the code PHOENIX and grids
of synthetic fluxes\footnote{The grids of synthetic spectra are available at http://phoenix.astro.physik.uni-goettingen.de/} simulated by \citet{Husser+13} for models with different metalicities and abundances of the $\alpha$-elements (O, Ne, Mg, Si, S, Ar, Ca, and Ti) employing version 16\footnote{Compared to the previous version, version 16 of the \textsc{phoenix} code employs an updated list of atomic and molecular lines and uses a new equation of state to simulate a model of the stellar atmosphere and the synthetic emerging flux \citep{Husser+13} of the {\sc phoenix} code} of the \textsc{phoenix} code \citep{H+B+99}. These grids were computed with a spectral resolution of R = 500000, which we reduced to R = 50000 during their transformation to the format read by the \textsc{fitsb2} code \citep{Khalack+17}. This latter resolution is close to the spectral resolution of FEROS (see Section~\ref{obs}).
%\citet{Husser+13} have calculated the grids of synthetic fluxes with spectral resolution R=500000, which we reduced to R=50000 to the format read by the \textsc{fitsb2} code . %The resolution used in simulations is close to the spectral resolution of FEROS (see Section~\ref{obs}).
%Comparing to the {\sc phoenix15} code, its new version, 16, employs an updated list of atomic and molecular lines and uses a new equation of state to simulate a model of stellar atmosphere and respective synthetic flux \citep{Husser+13}.

Two available FEROS spectra of HD~27463 were analysed individually and resulted in similar values of effective temperature, surface gravity and metallicity.
The best fit results obtained for one spectrum of HD~27463 that corresponds to the lowest $\chi^2$ %$\chi^2/_\nu$
are shown in Table~\ref{tab1} and Fig.~\ref{fig3}.
%The residuals between the synthetic and observed spectra shown in the Fig.~\ref{fig3}
This figure shows that the obtained best fit approximates the analysed Balmer line profiles relatively well even taking into account the contamination of Balmer wings by some metal lines.
%the two available spectra (see Table~\ref{tab1}) of In all cases
The fitting procedure has been carried out for different values of rotational velocity by including rotational broadening to the theoretical line profiles
%that was applied to all theoretical fluxes to widen them correspondingly
before the fitting of Balmer line profiles. In this case, the best fit value of v$\sin{i}$ is partially constrained by the metal lines present in the Balmer line wings. This approach allowed us to find the value that minimizes the discrepancy between the observed and simulated Balmer line profiles for v$\sin{i}$ = 27 $\pm$ 2 km s$^{-1}$ (see Table~\ref{tab1}). The derived value of v$\sin{i}$ has been confirmed by analysis of several Fe\,{\sc ii} line profiles visible in the two FEROS spectra of this star.

\begin{figure}
%\vspace{-110px}
%\hspace{30px}
%\includegraphics[width=3.5in]{HD27463_HR+TIC.eps}
%\includegraphics[width=3.5in]{HR_new_fit.eps}
\includegraphics[width=3.5in]{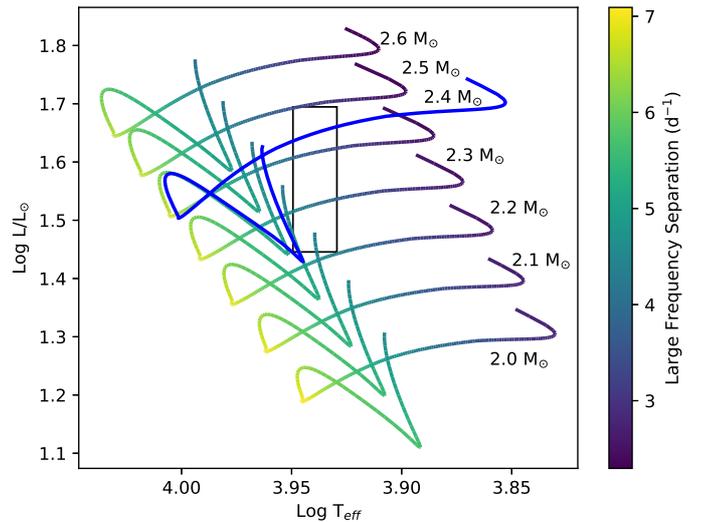}
%\vspace{110px}
\caption{\label{fig:HR}The main sequence evolution tracks for {\sc MESA} models from 2.0 to 2.6 M$_{\odot}$, all with zero convective overshoot and a metallicity derived from spectroscopy of HD~27463, are shown in %black.
colours that correspond to expected values of the large frequency separation.
The evolution track for the best fitting model with $M_{\rm \star}$ = 2.4 M$_{\odot}$, $f_{ov}$ = 0.014 %, and a ZAMS rotational velocity of {\bf 47} km s$^{-1}$
is shown in blue. The box indicates the $1-\sigma$ spectroscopic uncertainty box in the observed location of HD 27463.  }
%\caption{\label{fig:HR}The evolution tracks for {\sc mesa} models from 1.9 to 2.4 M$_{\odot}$, all with zero convective overshoot. The approximate location of HD~27463 is indicated with a star.}
\end{figure}

% assuming a rotational velocity v$\sin{i}$ = 27 km s$^{-1}$ derived for HD~27463 from the preliminary analysis of Si\,{\sc ii} line profiles. For the three available spectra the fitting procedure that employs the same type of grids of synthetic fluxes (for example, calculated with the PHOENIX15 code) results in almost the same values of $T_{\rm eff}$, $\log{g}$ and metallicity (see Table~\ref{tab2}). The effective temperatures provided by the use of the PHOENIX16 grids of synthetic fluxes \citep{Husser+13} appear to be always significantly smaller that the $T_{\rm eff}$ obtained employing the PHOENIX15 grids of synthetic fluxes \citep{Khalack+LeBlanc15a}. Meanwhile, both types of grids result in the same values of the surface gravity and metallicity and in the similar values of the radial velocity. Fig.~\ref{fig1} presents an example of the best fit of Balmer lines profiles obtained using the grids of synthetic fluxes simulated with the PHOENIX15 code \citep{Khalack+LeBlanc15a}.

\begin{figure}
%\vspace{-110px}
%\hspace{30px}
\includegraphics[width=3.5in]{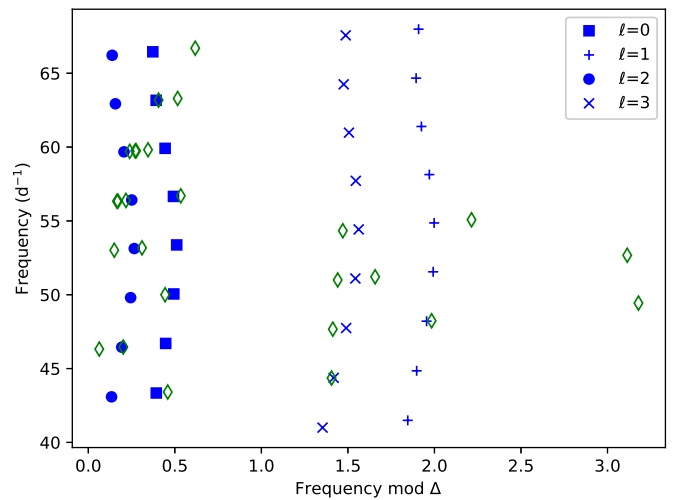}
%\vspace{110px}
\caption{\label{fig:bestfit}The echelle diagram for the best fitting model and the observed frequencies, plotted using the inferred large frequency separation of $\Delta \nu= 3.304$~d$^{-1}$.  The blue points show the model frequencies for $\ell = 0$ (squares), $\ell = 1$ (pluses), $\ell = 2$ (circles), and $\ell$ = 3 (crosses), while the green open diamonds specify the data derived from observations. The  model frequencies plotted here correspond to radial orders of $n = 11 - 19$.}
\end{figure}

%\subsection{Stellar Models}
\subsection{Analysis of stellar pulsations}
\label{stellar_model}

%Patrick Lenz:
%It would be nice to examine the reason for the sensitivity of the large frequency separation to convective overshooting for that specific range of radial orders (i.e., n slightly above 10) in more detail. This seems to be an interesting result.

%$-----------------------

We have calculated a grid of stellar structure and evolution models using {\sc mesa} version 11554 \citep{Paxton+11, Paxton+13, Paxton+15, Paxton+18, Paxton+19}, including the effects of convective overshoot. Model masses ranged from 2.0 to 2.6~M$_{\odot}$ in 0.1~M$_{\odot}$ increments, with rotation rates on the ZAMS between 35 and 100 km s$^{-1}$ taking into account the derived value of the equatorial velocity (see Table~\ref{tab1}). %Based on the spectral derivation of metallicity described above,
We used a scaled solar metallicity of Z = 0.0244 for all models to account for the inferred spectroscopic metallicity of the star.
Inside the convective core, convective mixing is calculated using mixing length theory \citep{bohm-vitense} %\citep[][MLT,]{bohm-vitense}
with a mixing length $\alpha$ = 1.73 (see e.g. \citealt{Montalban+04}).  Convective overshoot was included above the core using the diffusive exponential overshoot model described by  \citet{Herwig2000}:
%Convective overshoot was included above the core, and both above and below convective regions in the envelope. In {\sc mesa}, convection is treated as a diffusive process. The overshoot is determined using an exponential model \citep{Herwig2000}:
\begin{equation}
D_{ov} = D_0\exp\left(\frac{-2r}{f_{ov}H_P}\right),
\end{equation}
where $D_0$ is the diffusion coefficient at the convective boundary, $r$ is the radial distance from the core boundary, and $H_P$ is the pressure scale height at the core boundary.  The amount of overshoot is defined in terms of a fraction of the pressure scale height, $H_P$.  In our models, the amount of overshoot ranged from $f_{ov}$ = 0 to 0.04 in increments of 0.01.  As many of our best fitting models were found to have overshoot between 0.01 and 0.03, this region of the grid was subsequently refined with steps of 0.002.  The models were evolved from the pre-main sequence through the end of the main sequence.  Detailed models were saved at least every 20 time steps, and more frequently during certain evolutionary phases. The time steps were variable, and determined by the requirement that the relative change in the stellar structure be less than $10^{-4}$ from one time step to the next.
The evolution tracks for the zero-overshoot models are shown in Fig.~\ref{fig:HR}.
%Provide description of colours and error-box here.

We used {\sc gyre} \citep{GYRE} to calculate linear adiabatic pulsation frequencies for each main sequence model with $3.92 < \log{T_{\rm eff}} < 3.95$, consistent with both the temperature from the TIC and that derived from the Balmer lines in this work. We calculated model frequencies between 40 and 70~d$^{-1}$ for $\ell$ = 0, 1, 2, and 3. Initially, we calculated frequencies for $m=0$ modes only.  Once we had determined a small set of models which were good fits to the observations, we calculated the %$\ell=2,\
$m\neq0$ modes for comparison to the observed rotational splitting.
%Since HD~27463 is a relatively slow rotator, we restricted our calculations to $m = 0$ modes. For each model, we then calculated the large separation (i.e. the frequency difference between modes of the same angular degree, $\ell$, and consecutive radial order, $n$) for comparison to the observations.

\begin{figure}
%\vspace{-110px}
%\hspace{40px}
\includegraphics[width=3.5in]{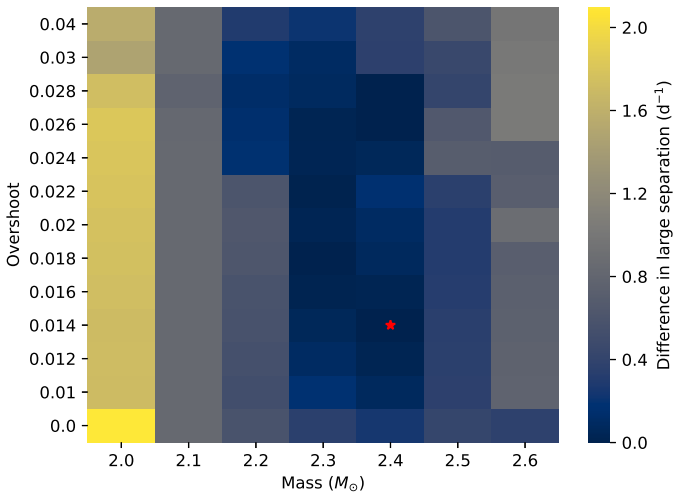}
%\vspace{110px}
\caption{\label{fig:DnuDistribution}The absolute value of the difference in the large frequency separation for the models in our grid compared to the observed value, shown as a function of mass and overshoot. Based on the difference in the large frequency separation and comparison to echelle diagrams, the best fitting model (red star) was determined to have $M_{\rm \star}$ = 2.4~M$_{\odot}$ and $f_{ov} = 0.014$.  }
\end{figure}

\begin{figure}
%\vspace{-110px}
%\hspace{40px}
\includegraphics[width=3.5in]{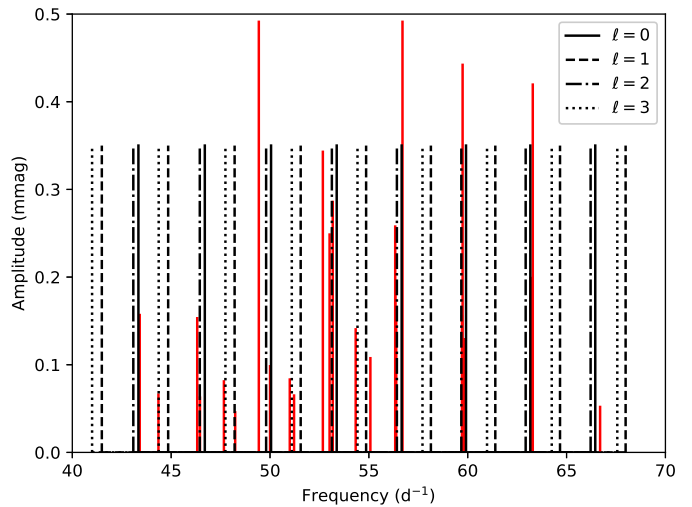}
%\vspace{110px}
\caption{\label{fig:individual} Individual frequency matches for the best fitting model. The red lines are the observed frequencies plotted versus amplitude, while the black lines are the $\ell = 0$ (solid), $\ell=1$ (dashed), $\ell=2$ (dot-dashed), and $\ell=3$ (dotted) model frequencies.}
\end{figure}

We used the 26 frequencies from Table~\ref{tab2_All}, all of which are greater than 43 d$^{-1}$ and appear to correspond to stellar pulsation modes to determine the large frequency separation
%which {\bf correspond to stellar pulsations and appear to be} greater than 43~d$^{-1}$ to determine large separation, defined as
%not 29 frequencies
\begin{equation}
\Delta \nu_0 = <\nu_{n+1,\ell} - \nu_{n,\ell} > .
\end{equation}
%as shown in Figure \ref{fig:echelle}.
%NOTE: Based on the SNR you sent me, I dropped frequencies 32, 34, and 35 from the list of frequencies I used in the fitting process.  They're probably too low amplitude to be significant, so you should remove them from Table 1.
%Removed 53.041, 53.362 from table1. Which is the 3rd one, 40.222?
Given the spectroscopically-determined $T_{\rm eff}$ value used as an input constraint to the modelling,
the best fitting large frequency separation is approximately 3.3~d$^{-1}$, which produces two clear ridges in the echelle diagram, as shown in Fig.~\ref{fig:bestfit}.
We then compared this derived large frequency separation to the values calculated for the models in our grid. The absolute value of the difference between the model and the observed large frequency separation is shown in Fig.~\ref{fig:DnuDistribution}. A number of models with masses between 2.2 and 2.5~M$_{\odot}$ have reasonably good fits to the large frequency separation. %, however, many of these models proved to not be good matches to the echelle diagram. %The difference in the model large separation and the observed large separation are shown as a function of mass and overshoot in Fig.~\ref{fig:DnuDistribution}.
For models with the smallest difference in the large frequency separation, the echelle diagrams were plotted.
Not all models with a large frequency separation of 3.3 d$^{-1}$ were good fits to the observed echelle diagram, so we also compared individual frequency fits, as shown in Fig.~\ref{fig:individual}.

%It is known that forward asteroseismic modelling of pulsating stars includes a correlation between mass and overshoot (see e.g. \citep{Moravveji+16,Moravveji+15,Buysschaert+18,Aerts+18a}).
%%There appears to be a slight correlation between mass and overshoot, as models with higher mass also have higher overshoot.
%Models with low mass and high overshoot are not good fits to the observed large frequency separation in HD~27463, neither are models with the highest mass and lowest overshoots.

%\begin{figure}
%\vspace{-110px}
%\hspace{30px}
%\includegraphics[width=3.5in]{ObservedEchelle.eps}
%\vspace{110px}
%\caption{\label{fig:Dnu}The echelle diagram based on 29 frequencies extracted from the \textit{TESS} data.  The large separation used in this plot is 3.28 d$^{-1}$.}
%\end{figure}

%!!!New paragraph here!!! Talk about Bedding et al. (2019)
HD~27463 has been analysed recently by \citet{Bedding+19}
%assuming that it is a young intermediate-mass $\delta$~Scuti star. Under this assumption, the aforementioned authors
who have performed mode identification using the echelle diagram and obtained a large frequency separation of 6.91 d$^{-1}$, more than twice our result.
Our models show that HD~27463 is an evolved $\delta$~Scuti star with an age of 5 $\times 10^8$~yr and an effective temperature which is similar to the one found from the analysis of Balmer line profiles (see Table~\ref{tab1}). In Fig.~\ref{fig:HR}, the $1-\sigma$ spectroscopic uncertainty box that specifies the observed location of HD~27463 on the HR diagram is constructed taking into account uncertainties provided by the TIC for the estimates of stellar luminosity and effective temperature \citep{Stassun+18,Stassun+19}. Fig.~\ref{fig:HR} clearly demonstrates that the large frequency separation around $\Delta \nu_0= 6.9$~d$^{-1}$ is a characteristic of very young stars %at this metallicity
(see also \citealt{Bedding+19}).

We have repeated our fitting procedure assuming a large frequency separation of 6.91~d$^{-1}$ derived by \citet{Bedding+19}. In this case, we found that the best fitting model has a large frequency separation of 6.707~d$^{-1}$ and is a 2.0 M$_{\odot}$ model with a core overshoot of 0.03 and a ZAMS equatorial rotation velocity of 40 km s$^{-1}$. The echelle diagram and frequency fit for this model are shown in Figs.~\ref{fig:largeA} and~\ref{fig:individual6} respectively. The model corresponds to a young star that is effectively on the ZAMS. %This model fails to explain a number of observed modes, and
The derived $\chi^2$ for the best frequency fit is an order of magnitude larger than our best fit with $\Delta \nu_0$ = 3.3~d$^{-1}$, which takes into account the spectroscopic constraints on temperature and metallicity (see Figs.~\ref{fig:HR},~\ref{fig:bestfit},~\ref{fig:DnuDistribution} and ~\ref{fig:individual}). This finding clearly indicates that the large frequency separation is not a unique factor on which we should rely to determine the fundamental stellar parameters. The spectroscopy and specifically spectroscopic estimates of $T_{\rm eff}$ and metallicity are needed to break degeneracies amongst fundamental parameters when performing forward seismic modelling (see \citealt{Aerts+18a, Buysschaert+18}).
%. It also highlights the importance of spectroscopic observations to complement the photometry.
%In this case, we find the best fitting model is a 2.0 $M_{\odot}$ model with a core overshoot of 0.03 and a ZAMS equatorial velocity of 40 km s$^{-1}$.  The model is extremely young, and is effectively on the ZAMS. The best fitting echelle diagram and the resulting frequency fits are shown in Appendix~\ref{append}.
%This model has a large separation of 6.707~d$^{-1}$.  The resulting frequency fits are also shown in Appendix A have a $\chi^2$ an order of magnitude larger than our best fitting model, which takes into account spectroscopic constraints on temperature and metallicity. This indicates that the large separation is not a unique determiner of the stellar parameters, since spectroscopy and specifically estimates of $T_{\rm eff}$ and metallicity are needed to break degeneracies amongst fundamental parameters when performing forward seismic modelling (see \citealt{Aerts+18a, Buysschaert+18}).

Based on our simulations, the
%this comparison, our
best fitting model was determined to have
%Our best fitting model was determined to have an
%an overshoot $f_{ov} = 0.014$, which gives
a large frequency separation of 3.304 d$^{-1}$,
and is compared to the observations in Fig.~\ref{fig:bestfit}. This model has a mass of 2.4~M$_{\odot}$, a ZAMS rotation velocity of 47 $\pm$ 5 km s$^{-1}$, and an overshoot of $f_{ov} = 0.014$.  %The effective temperature is 8770~K, in good agreement with the spectroscopic measurements.  The luminosity is 45.2 $L_{\odot}$, and $\log g = 3.887$.
The derived values of effective temperature, $T_{\rm eff} = 8800 \pm 100$~K, and surface gravity, $\log g = 3.89 \pm 0.05$, are in good agreement with the spectroscopic measurements (see Table~\ref{tab1}). The stellar luminosity of 42.5~L$_{\odot}$ derived from the best fitting model is comparable to the value reported in the TIC.

%Individual frequency matches for the best fitting model. The red lines are the observed frequencies plotted versus amplitude, while the assorted black lines are the model frequencies, plotted versus $\ell$ on the right hand axis.

\begin{figure}
%\vspace{-110px}
%\hspace{30px}
\includegraphics[width=3.5in]{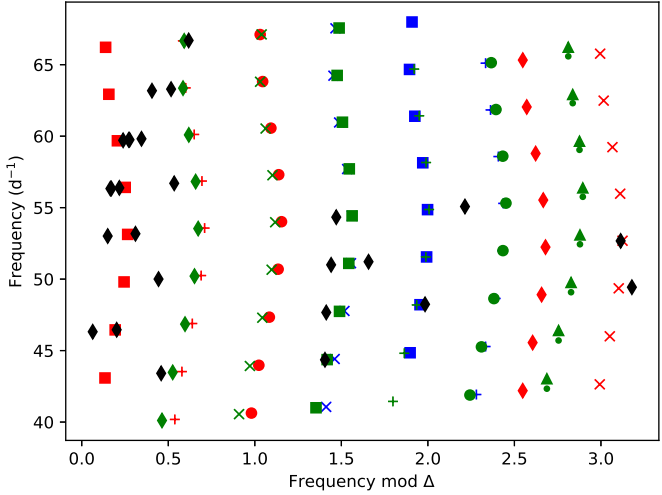}
%\vspace{110px}
\caption{\label{fig:splitting} The mode splitting in the $\ell$ = 1 (blue), 2 (red), 3 (green) modes for the best fitting model. The $m$ = -3, -2, -1, 0, 1, 2 and 3 frequencies are denoted by triangles, diamonds, crosses,
%$\times$,
squares, pluses, circles, and dots respectively. The black diamonds indicate the observed frequencies. The range of radial orders shown is the same as in Fig.~\ref{fig:bestfit}. }
\end{figure}

During the simulation of stellar structure, the core hydrogen fraction evolves with the star and is one of the output parameters provided by the code {\sc mesa} \citep{Paxton+11, Paxton+13, Paxton+15, Paxton+18, Paxton+19}. Assuming that initial core hydrogen fraction is $X_{\rm c}$=0.7, our best fitting model was determined to have an age of 5 $\times 10^8$ yr, which corresponds to $X_{\rm c}$=0.33. These parameters are derived from the best fitting model to the echelle diagram %{\bf taking into account the rotational splitting of pulsation modes}
(see Fig.~\ref{fig:splitting}).

The amount of overshooting we found in this model is similar to that found for KIC~10526294 \citep{Moravveji+15}, and slightly lower than that found for KIC~7760680 \citep{Moravveji+16}.  However, both of these stars are late-type B stars, and show $g$ mode pulsations typical of Slowly Pulsating B (SPB) stars.  Similar studies of a magnetic B-type star (HD~43317) found a much lower convective overshoot ($f_{ov} = 0.004$) and the authors conclude that convective mixing is suppressed by the magnetic field \citep{Buysschaert+18}.
%Do we know the magnetic field strength of HD27463?  How does it compare to this star?
%Magnetic field should be no more that few hundreds Gauss, because the spectral line profiles are not split.
More recently, \citet{Mombarg+19}
%Mombarg et al. (2019)
performed ensemble asteroseismic modelling of gravity modes in 37 $\gamma$~Doradus stars observed by the Kepler Space mission. %Mombarg et al. (2019)
\citet{Mombarg+19} found best-fitting diffusive exponential overshooting values that range between 0 and 0.03 for these main-sequence intermediate-mass stars.

Once a best fitting model was established,
%we had established a best fitting model,
we calculated rotational splitting in the $\ell$= 1, 2 and 3 modes to see if this provided matches for some of the unmatched frequencies in the echelle diagram, since rotation also has a significant impact on pulsation mode frequencies \citep{Aerts+10}.
%we calculated the mode splitting for the $\ell = 2$ modes in this model.
As shown in Fig.~\ref{fig:splitting}, the observed mode splitting agrees well with our model calculations.
For a model that is initially rotating at 100~km~s$^{-1}$, the present-day rotation rate has dropped to 47~km~s$^{-1}$ ($\Omega\Big/\Omega_{crit}$ = 0.12) at $X_{\rm c}$ = 0.33.
%Based on these calculations, we present our best fitting mode identification in Table~\ref{tab2_All}.

%For models with the smallest difference in the large separation, the echelle diagrams were plotted by eye. Our best fitting model was determined to have a mass of 2.3~M$_{\odot}$ and an overshoot of $f_{ov} = 0.026$, and is compared to the observations in Figure \ref{fig:bestfit}.
%an overshoot of ?3.3304?, This model has

%The model was determined to have an age of 6.007 $\times 10^8$ years, which corresponds to a core hydrogen fraction of 0.353.
%The final echelle diagram is shown in Figure \ref{fig:bestfit}.

%\bibliographystyle{yahapj}
%\bibliography{writeup}

\begin{thebibliography}{99}
% When the are 8 and less authors -- all their names should be given.
% Only when their number is bigger than 8, use "et al."

%\bibitem[\protect\citeauthoryear{Abt \& Morell}{1995}]{Abt+Morell95} Abt H.A., Morell N.I., 1995, ApJS, 99, 135

%\bibitem[\protect\citeauthoryear{Adelman}{1987}]{Adelman87} Adelman S.J., 1987, A\&AS, 67, 353

\bibitem[\protect\citeauthoryear{Aerts et al.}{2018}]{Aerts+18a} Aerts C., Molenberghs G., Michielsen M., Pedersen M.G., Bj{\"o}rklund R., Johnston C., Mombarg J.S.G., Bowman D.M., et al. 2018, \apjs, 237, 15

\bibitem[\protect\citeauthoryear{Aerts et al.}{2010}]{Aerts+10} Aerts C., Christensen-Dalsgaard J., Kurtz D.~W., 2010, Asteroseismology, Springer


\bibitem[\protect\citeauthoryear{Alecian \& Stift}{2010}]{Alecian+Stift10} Alecian G., Stift M.J., 2010, A\&A, 516, 53

\bibitem[\protect\citeauthoryear{Antoci et al.}{2014}]{Antoci+14} Antoci V., Cunha M., Houdek G., Kjeldsen H., Trampedach R., Handler G., L{\"u}ftinger T., Arentoft T., et al. 2014, ApJ, 796, 118
%Murphy S.

\bibitem[\protect\citeauthoryear{Babcock}{1958}]{Babcock58} Babcock H.W., 1958, ApJ, 128, 228

\bibitem[\protect\citeauthoryear{Baize \& Petit}{1989}]{Baize+Petit89} Baize P., Petit M., 1989, A\&AS, 77, 497

\bibitem[\protect\citeauthoryear{Balona et al.}{2019}]{Balona+19} Balona L.A., Holdsworth D.L., Cunha M.S., 2019, MNRAS, tmp.1356

\bibitem[\protect\citeauthoryear{Bedding et al.}{2019}]{Bedding+19} Bedding T.R., Murphy S.J., Hey D.R., Huber D., Li T., Li G., Li Y., Smalley B., et al. 2019, submitted

%\bibitem[\protect\citeauthoryear{Bevington}{1969}]{Bevington+69} Bevington P.R., 1969, The data reduction and error analysis for the physical sciences, McGraw-Hill (New Yourk)

%\bibitem[\protect\citeauthoryear{Bidelman}{1988}]{Bidelman88} Bidelman W.P., 1988, PASP, 100, 1084

\bibitem[\protect\citeauthoryear{Bigot \& Kurtz}{2011}]{Bigot+Kurtz11} Bigot L., Kurtz D.W., 2011, A\&A, 536, 73

\bibitem[\protect\citeauthoryear{B{\"o}hm-Vitense}{1958}]{bohm-vitense} B{\"o}hm-Vitense E., 1958, Zeitschrift fur Astrophysik, 46, 108

\bibitem[\protect\citeauthoryear{Bowman \& Kurtz}{2018}]{Bowman+Kurtz18a} Bowman D.M., Kurtz D.W., 2018, MNRAS, 476, 3169

\bibitem[\protect\citeauthoryear{Bowman et al.}{2018}]{Bowman+18b} Bowman D.M., Buysschaert B., Neiner C., P{\'a}pics P.I., Oksala M.E., Aerts C., 2018, \aap, 616, A77

\bibitem[\protect\citeauthoryear{Bowman}{2017}]{Bowman16} Bowman D.M., 2017, Amplitude Modulation of Pulsation Modes in Delta Scuti Stars, Springer Theses series, Springer
%PhD thesis, Jeremiah Horroks Institute, University of Central Lancashire, UK

\bibitem[\protect\citeauthoryear{Bowman et al.}{2016}]{Bowman+16} Bowman D.M., Kurtz D.W., Breger M., Murphy S.J., Holdsworth D.L., 2016, MNRAS, 460, 1970

\bibitem[\protect\citeauthoryear{Breger}{2000}]{Breger00} Breger M., 2000, MNRAS, 313, 129

%\bibitem[\protect\citeauthoryear{Bressan et al.}{2012}]{Bressan+12} Bressan M.P., Girardi L., Salasnich B. et al., 2012, MNRAS, 427, 127
%; Dal Cero, Claudia; Rubele, Stefano; Nanni, Ambra

\bibitem[\protect\citeauthoryear{Buysschaert et al.}{2018}]{Buysschaert+18} Buysschaert B., Aerts C., Bowman D.M., Johnston C., Van Reeth T., Pedersen M.G., Mathis S., Neiner C., 2018, \aap, 616, A148

%\bibitem[\protect\citeauthoryear{Bychkov et al.}{2003}]{Bychkov+03} Bychkov V. D., Bychkova L. V., Madej J., 2003, A\&A, 407, 631

\bibitem[\protect\citeauthoryear{Campante et al.}{2016}]{Campante+16} Campante T.L., Schofield M., Kuszlewicz J.S., et al., 2016, ApJ, 830, 138

%\bibitem[\protect\citeauthoryear{Carrier et al.}{2002}]{Carrier+02}  Carrier F., North P., Udry S., Babel J., 2002, A\&A, 394, 151

%\bibitem[\protect\citeauthoryear{Castelli}{2005}]{Castelli05} Castelli F., 2005, Memorie della Societa Astronomica Italiana Supplementi, 8, 44

%\bibitem[\protect\citeauthoryear{Castelli et al.}{2017}]{Castelli+17} Castelli F., Cowley C.R., Ayers T.R., et al., 2017, A\&A, accepted
%Catanzaro G., Leone F }

\bibitem[\protect\citeauthoryear{Cleveland}{1979}]{Cleveland79} Cleveland W.S., 1979, Journal of the American Statistical Association, 74, N368, 829

\bibitem[\protect\citeauthoryear{Cunha et al.}{2019}]{Cunha+19} Cunha M.S., Antoci V., Holdsworth D.L., Kurtz D.W., Balona L. A., Bogn\'{a}r Zs., Bowman D.M. , Guo Z., et al., 2019, MNRAS, 487, 3523

%\bibitem[\protect\citeauthoryear{Cowley et al.}{1969}]{Cowley+69} Cowley A., Cowley C., Jaschek M., Jaschek C., 1969, AJ, 74, 375

%\bibitem[\protect\citeauthoryear{Cowley et al.}{1978}]{Cowley+78} Cowley C.R., Elste G.H., Urbanski J.L., 1978, PASP, 90, 536

\bibitem[\protect\citeauthoryear{David-Uraz et al.}{2019}]{David-Uraz+19} David-Uraz A., Neiner C., Sikora J., Bowman D.M., Petit V., Chowdhury S., Handler G., Pergeorelis M., et al.,  2019, MNRAS, 487, 304

%\bibitem[\protect\citeauthoryear{Donati et al.}{1997}]{Donati+97} Donati J.-F., Semel M., Carter B.D., Rees D.E., Cameron A.C., 1997, MNRAS, 291, 658

%\bibitem[\protect\citeauthoryear{Donati et al.}{2006}]{Donati+06} Donati J.-F., Catala C., Landstreet J.D., Petit P., 2006, in Casini R. \& Lites B.W., eds, ASP Conf. Ser.
%Vol. 358, Solar Polarization 4, Boulder, p. 362

\bibitem[\protect\citeauthoryear{Dupret et al.}{2019}]{Dupret+04} Dupret M.A., Grigahc{\`e}ne A., Garrido R., Gabriel M., Scuflaire R., 2004, \aap, 414, L17

%\bibitem[\protect\citeauthoryear{Erspamer \& North}{2003}]{Erspamer+North03} Erspamer D., North P., 2003, A\&A, 398, 1121

\bibitem[\protect\citeauthoryear{ESA}{1997}]{ESA97} ESA, 1997, The Hipparcos and Tycho Catalogues (ESA, SP Series 1200, Noordwijk: ESA)

%\bibitem[\protect\citeauthoryear{Green et al.}{2014}]{Green+14} Green G.M., Schlafly E.F.; Finkbeiner D.P. et al., 2014, ApJ, 783, 114

%\bibitem[\protect\citeauthoryear{Green et al.}{2015}]{Green+15} Green G.M., Schlafly E.F.; Finkbeiner D.P. et al., 2015, ApJ, 810, 25

\bibitem[\protect\citeauthoryear{Gontcharov}{2006}]{Gontcharov2006} Gontcharov G.A., 2006, PAZh, 32, 844

%\bibitem[\protect\citeauthoryear{Grevesse et al.}{2010}]{Grevesse+10} Grevesse N., Asplund M., Suaval A.J., Scott P., 2010, Ap\&SS, 328, 179

%\bibitem[\protect\citeauthoryear{Grevesse et al.}{2015}]{Grevesse+15} Grevesse N., Scott P., Asplund M., Sauval A.J., 2015, A\&A, 573, 27

%\bibitem[\protect\citeauthoryear{Hauck \& Mermilliod}{1998}]{Hauck+Mermilliod98} Hauck B., Mermilliod M., 1998,  A\&ASS, 129, 431

\bibitem[\protect\citeauthoryear{Hartkopf et al.}{2012}]{Hartkopf+12} Hartkopf W.~I., Tokovinin A., Mason B.~D., 2012, \aj, 143, 42

\bibitem[\protect\citeauthoryear{Hauschildt et al.}{1997}]{Hauschildt+97} Hauschildt P. H., Baron E., Allard F., 1997, ApJ, 483, 390

\bibitem[\protect\citeauthoryear{Hauschildt \& Baron}{1999}]{H+B+99}Hauschildt P. H., Baron E., 1999, JCoAM, 109, 41

\bibitem[\protect\citeauthoryear{Herwig}{2000}]{Herwig2000} Herwig F., 2000, \aap 360, 952

\bibitem[\protect\citeauthoryear{Holdsworth et al.}{2018}]{Holdsworth+18} Holdsworth D.L., Cunha M.S., Shibahashi H., Kurtz D.W., Bowman D.M., 2018, MNRAS, 480, 2976

\bibitem[\protect\citeauthoryear{Holdsworth et al.}{2016}]{Holdsworth+16} Holdsworth D.L., Kurtz D.W., Smalley B., Saio H., Handler G., Murphy S.J., Lehmann H., 2016, MNRAS, 462, 876

\bibitem[\protect\citeauthoryear{Holdsworth et al.}{2014}]{Holdsworth+14} Holdsworth D.L., Smalley B., Gillon M., Clubb K.I., Southworth J., Maxted P.F.L., Anderson D.R., Barros S.C.C, et al., 2014, MNRAS, 439, 2078

\bibitem[\protect\citeauthoryear{Houk \& Cowley}{1975}]{Houk+Cowley+75} Houk N., Cowley A.P., 1975, Michigan Spectral Survey, 1, 0

%\bibitem[\protect\citeauthoryear{Hui-Bon-Hoa et al.}{2000}]{HBH+00} Hui-Bon-Hoa A., LeBlanc F., Hauschildt P. H., 2000, ApJ, 535, L43

\bibitem[\protect\citeauthoryear{Husser et al.}{2013}]{Husser+13} Husser T.-O., Wende-von Berg S., Dreizler S., Homeier D., Reiners A., Barman T., Hauschildt P. H.,
Husser T.-O., et al., 2013, A\&A, 553, 6
%Homeier, D.; Reiners, A.; Barman, T.; Hauschildt, P. H.
%Husser T.-O., Kamann S., Dreizler S., Hauschildt P. H.2012, in: Prugniel P. \& Harinder P., eds, Astronomical Society of India Conference Series, 6, 71

\bibitem[\protect\citeauthoryear{Jenkins et al.}{2016}]{Jenkins+16} Jenkins 2016, Proc. SPIE 9913, Software and Cyberinfrastructure for Astronomy IV, 99133E
%A description of the SPOC pipeline can be found here: Jenkins et al., 2016. A PDF version of the article can be downloaded here provided acknowledgement of the full SPIE citation: Jenkins et al. 2016. "The TESS science processing operations center", Proc. SPIE 9913, Software and Cyberinfrastructure for Astronomy IV, 99133E (August 8, 2016); doi:10.1117/12.2233418.

\bibitem[\protect\citeauthoryear{Joshi et al.}{2016}]{Joshi+16} Joshi S., Martinez P., Chowdhury S., et al., 2016, A\&A, 590, A116

\bibitem[\protect\citeauthoryear{Kaufer et al.}{1999}]{Kaufer+99} Kaufer A., Stahl O., Tubbesing S., N{\o}rregaard P., Avila G., Francois P.,Pasquini L., Pizzella A., 1999, Messenger, 95, 8

%\bibitem[\protect\citeauthoryear{Khalack }{2018}]{Khalack18} Khalack V., 2018, MNRAS, 477, 882

%\bibitem[\protect\citeauthoryear{Khalack et al.}{2017}]{Khalack+17} Khalack V., Gallant G., Thibeault C., 2017, MNRAS, 471, 926

%\bibitem[\protect\citeauthoryear{Khalack \& Landstreet}{2012}]{Khalack+Landstreet12} Khalack V., Landstreet J., 2012, MNRAS, 427, 569

\bibitem[\protect\citeauthoryear{Khalack \& LeBlanc}{2015a}]{Khalack+LeBlanc15a} Khalack V., LeBlanc F., 2015a, AJ, 150, 1, id.2

\bibitem[\protect\citeauthoryear{Khalack \& LeBlanc}{2015b}]{Khalack+LeBlanc15b} Khalack V., LeBlanc F., 2015b, Advances in Astronomy and Space Physics, 5, 3

%\bibitem[\protect\citeauthoryear{Khalack \& Wade}{2006}]{Khalack+Wade06} Khalack V., Wade G., 2006, A\&A, 450, 1157

%\bibitem[\protect\citeauthoryear{Khalack et al.}{2007}]{Khalack+07} Khalack V., LeBlanc F., Bohlender D., Wade G., Behr B. B., 2007, A\&A, 466, 667

%\bibitem[\protect\citeauthoryear{Khalack et al.}{2008}]{Khalack+08} Khalack V., LeBlanc F., Behr B.B., Wade G.A., Bohlender D., 2008, A\&A, 477, 641

%\bibitem[\protect\citeauthoryear{Khalack et al.}{2010}]{Khalack+10} Khalack V., LeBlanc F., Behr B.B., 2010, MNRAS, 407, 1767

%\bibitem[\protect\citeauthoryear{Khalack et al.}{2013}]{Khalack+13} Khalack V., Yameogo B., Thibeault C., LeBlanc F., 2013, in Petit P., Jardine M. \& Spruit H.C., eds, Proc. IAU Symp. 302, Magnetic Fields throughout Stellar Evolution. Cambridge univ. press, Cambridge,  p. 272

%\bibitem[\protect\citeauthoryear{Khalack et al.}{2014}]{Khalack+14} Khalack V., Yameogo B., LeBlanc F. et al., 2014, MNRAS, 445, 4086
%Fontaine G., Green E. , Van Grootel V., Petit P.

%\bibitem[\protect\citeauthoryear{Khalack \& Poitras}{2015}]{Khalack+Poitras15} Khalack V., Poitras P., 2015, in Meynet G., Georgy C., Groh J.H. \& Stee Ph., eds, Proc. IAU Symp. 307, New Windows on MAssive Stars. Cambridge univ. press, Cambridge, p. 383

\bibitem[\protect\citeauthoryear{Khalack et al.}{2017}]{Khalack+17} Khalack V., Gallant G., Thibeault C., 2017, MNRAS, 471, 926

\bibitem[\protect\citeauthoryear{Kochukhov et al.}{2019}]{Kochukhov+19} Kochukhov O., Shultz M., Neiner C., 2019, A\&A, 621, 47K

%\bibitem[\protect\citeauthoryear{Khokhlova}{1975}]{Khokhlova75} Khokhlova V.L., 1975, Azh, 52, 950

%\bibitem[\protect\citeauthoryear{Kramida et al.}{2015}]{Kramida+15}Kramida, A., Ralchenko, Yu., Reader, J., and NIST ASD Team (2015). NIST Atomic Spectra Database (ver. 5.3). Available: http://physics.nist.gov/asd. National Institute of Standards and Technology, Gaithersburg, MD
%Kramida~A., Ralchenko~Yu., Reader~J., \& NIST ASD Team 2013, NIST Atomic Spectra Database (ver. 5.1). Available: http://physics.nist.gov/asd. National Institute of Standards and Technology, Gaithersburg, MD

%\bibitem[\protect\citeauthoryear{Kupka et al.}{2000}]{Kupka+00} Kupka F., Ryabchikova T.A., Piskunov N.E., Stempels H.C., Weiss W.W., 2000, Baltic Astronomy, 9, 590

\bibitem[\protect\citeauthoryear{Kurtz}{1978}]{Kurtz78} Kurtz D.W., 1978, IBVS, 1436, 1

\bibitem[\protect\citeauthoryear{Kurtz}{1982}]{Kurtz82}	Kurtz D.W., 1982, MNRAS, 200, 807

\bibitem[\protect\citeauthoryear{Kurtz et al.}{2006}]{Kurtz+06} Kurtz D.W., Elkin V.G., Cunha M.S., Mathys G., Hubrig S., Wolff B., Savanov I., 2006, MNRAS, 372, 286

%\bibitem[\protect\citeauthoryear{Landstreet}{1988}]{Landstreet88} Landstreet J. D., 1988, ApJ, 326, 967

%\bibitem[\protect\citeauthoryear{Landstreet et al.}{2008}]{Landstreet+08} Landstreet J. D., Silaj J., Andretta V. et al., 2008, A\&A, 481, 465

%\bibitem[\protect\citeauthoryear{Landstreet et al.}{2015}]{Landstreet+15} Landstreet J. D., Bagnulo S., Valyavin G.G. et al., 2015, A\&A, 580, 120

%\bibitem[\protect\citeauthoryear{LeBlanc et al.}{2009}]{LeBlanc+09} LeBlanc F., Monin D., Hui-Bon-Hoa A., Hauschildt P.H., 2009, A\&A, 495, 337

%\bibitem[\protect\citeauthoryear{LeBlanc et al.}{2010}]{LeBlanc+10} LeBlanc F., Hui-Bon-Hoa A., Khalack V., 2010, MNRAS, 409, 1606

%\bibitem[\protect\citeauthoryear{LeBlanc et al.}{2015}]{LeBlanc+15} LeBlanc F., Khalack V., Yameogo B., Thibeault C., Gallant I., 2015, MNRAS, 453, 3766

%\bibitem[\protect\citeauthoryear{van Leeuwen}{2007}]{vanLeeuwen07} van Leeuwen F., 2007, A\&A, 474, 653

\bibitem[\protect\citeauthoryear{Lenz \& Breger}{2005}]{Lenz+Breger05} Lenz P., Breger M., 2005, Commun. in Asteroseismology, 146, 53

\bibitem[\protect\citeauthoryear{L\"{u}ftinger et al.}{2010}]{Luftinger+10} L\"{u}ftinger T., Fr\"{o}hlich H.-E., Weiss W.W., Petit P., Auri\`{e}re M., Nesvacil N., Gruberbauer M., Shulyak D., et al. 2010, A\&A, 509, 43L

%\bibitem[\protect\citeauthoryear{Lyubimkov}{2007}]{Lyubimkov89} Lyubimkov L.S., 1989, Astrophysics 31, I3, 519

%\bibitem[\protect\citeauthoryear{Maitzen}{1984}]{Maitzen1984} Maitzen H.M., 1984, A\&A, 138, 493

%\bibitem[\protect\citeauthoryear{Makarov \& Kaplan}{2005}]{Makarov+Kaplan05} Makarov V.V., Kaplan G.H., 2005, AJ 129, 2420

\bibitem[\protect\citeauthoryear{Martinez \& Kurtz}{1994}]{Martinez+Kurtz94} Martinez P., Kurtz D.W., 1994, MNRAS, 271, 129

%\bibitem[\protect\citeauthoryear{Mashonkina et al.}{2005}]{Mashonkina+05} Mashonkina L., Ryabchikova T., Ryabtsev A., 2005, A\&A, 441, 309

%\bibitem[\protect\citeauthoryear{Mathys \& Hubrig}{1997}]{Mathys+Hubrig97} Mathys G., Hubrig S., 1997, A\&ASS, 124, 475

\bibitem[\protect\citeauthoryear{Mathys et al.}{2019}]{Mathys+19} Mathys G., Khalack V., Landstreet J., 2019, A\&A, submitted

\bibitem[\protect\citeauthoryear{Maury \& Pickering}{1897}]{Maury+Pickering97} Maury A.C., Pickering E.C., 1897, Annals of Harvard College Observatory, 28, 1

%\bibitem[\protect\citeauthoryear{Michaud}{1970}]{Michaud70} Michaud G., 1970, ApJ, 160, 641

%\bibitem[\protect\citeauthoryear{Michaud et al.}{1983}]{Michaud+83} Michaud G., Tarasick D., Charland Y., Pelletier C., 1983, ApJ, 269, 239

%\bibitem[\protect\citeauthoryear{Michaud et al.}{2015}]{Michaud+15} Michaud G., Alecian G., Richer J., 2015, Atomic Diffusion in Stars, Springer International Publishing, Switzerland, p.327

%\bibitem[\protect\citeauthoryear{Mihalas}{1970}]{Mihalas65} Mihalas D., 1965, ApJS, 9, 321

\bibitem[\protect\citeauthoryear{Mombarg et al.}{2019}]{Mombarg+19} Mombarg J.~S.~G., Van Reeth T., Pedersen M.~G., Molenberghs G., Bowman D.~M., Johnston C., Tkachenko A., Aerts C., 2019, MNRAS, 485, 3248

\bibitem[\protect\citeauthoryear{Montalb{\'a}n et al.}{2004}]{Montalban+04} Montalb{\'a}n J., D'Antona F., Kupka F., Heiter U., 2004, A\&A, 416, 1081

\bibitem[\protect\citeauthoryear{Montgomery \& O'Donoghue }{1999}]{Montgomery+Odonoghue99} Montgomery M.H., O'Donoghue D., 1999, Delta Scuti Star Newsletter, 13, 28

\bibitem[\protect\citeauthoryear{Moravveji et al.}{2016}]{Moravveji+16} Moravveji E., Townsend R.H.D., Aerts C., Mathis S., 2016, \apj, 823, 130

\bibitem[\protect\citeauthoryear{Moravveji et al.}{2015}]{Moravveji+15} Moravveji E., Aerts C., P{\'a}pics P.I., Triana S.A.,Vandoren B., 2015, \aap, 580, A27


%\bibitem[\protect\citeauthoryear{Morgan}{1932}]{Morgan32} Morgan W.W., 1932, ApJ, 75, 46

%\bibitem[\protect\citeauthoryear{Napiwotzki et al.}{1993}]{Napiwotzky+93} Napiwotzki R., Sch\"{o}nberner D., Wenske V., 1993, A\&A, 268, 653

\bibitem[\protect\citeauthoryear{Napiwotzki et al.}{2004}]{Napiwotzki+04} Napiwotzki R., Yungelson L., Nelemans G., Marsh T. R., Leibundgut B., Renzini R., Homeier D., Koester D., et al., 2004,
in: Hilditch R. W., Hensberge H. \& Pavlovski K., eds, ASP Conf. Ser. Vol. 318, Spectroscopically and Spatially Resolving the Components of the Close Binary Stars, San Francisco, p. 402

%\bibitem[\protect\citeauthoryear{Ndiaye et al.}{2017}]{Ndiaye+17} Ndiaye M., LeBlanc F., Khalack V., 2017, MNRAS, submitted

%\bibitem[\protect\citeauthoryear{Netopil et al.}{2008}]{Netopil+08} Netopil M., Paunzen E., Maitzen H.M., North P., Hubrig S., 2008, A\&A, 491, 545

\bibitem[\protect\citeauthoryear{Pamyatnykh}{2000}]{Pamyatnykh00} Pamyatnykh A.A., 2000, in: Breger M., Montgomery M., eds, Astronomical Society of the Pacific Conference Series Vol. 210, Delta Scuti and Related Stars, p. 215

\bibitem[\protect\citeauthoryear{Paxton et al.}{2019}]{Paxton+19} Paxton B., Smolec R., Schwab J., Gautschy A., Bildsten L., Cantiello M., Dotter A., Farmer R., et al. 2019, ApJS, 243, 10 %arXiv:1903.01426

\bibitem[\protect\citeauthoryear{Paxton et al.}{2018}]{Paxton+18} Paxton B., Schwab J., Bauer E.B., Bildsten L., Blinnikov S., Duffell P., Farmer R., Goldberg J.A., et al., 2018, ApJS, 234, 34

\bibitem[\protect\citeauthoryear{Paxton et al.}{2015}]{Paxton+15} Paxton B., Marchant P., Schwab J., Bauer E.B., Bildsten L., Cantiello M., Dessart L., Farmer R., et al., 2015, ApJS, 220, 15

\bibitem[\protect\citeauthoryear{Paxton et al.}{2013}]{Paxton+13} Paxton B., Cantiello M., Arras P., Bildsten L., Brown E.F., Dotter A., Mankovich C., Montgomery M.H., et al, 2013, ApJS, 208, 4

\bibitem[\protect\citeauthoryear{Paxton et al.}{2011}]{Paxton+11} Paxton B., Bildsten L., Dotter A., Herwig F., Lesaffre P., Timmes F., 2011, ApJS, 192, 3

%\bibitem[\protect\citeauthoryear{Pasinetti~Fracassini et al.}{2001}]{Pasinetti-Fracassini+01} Pasinetti~Fracassini L.E., Pastori L., Covino S., Pozzi A., 2001, A\&A 367, 521

%\bibitem[\protect\citeauthoryear{Plaskett et al.}{1922}]{Plaskett+22} Plaskett J.S., Harper W.E., Young R.K., Plaskett H.H., 1922, Publications of the Dominion Astrophys Obsesrvatory, 1, 287

%\bibitem[\protect\citeauthoryear{Press et al. }{1992}]{Press+92} Press W. H., Teukolsky S. A., Vetterling W. T., Flannery B. P., 1992, Numerical recipes in C: the
%art of scientific computing, 2nd ed., Cambridge University Press, 995p

%\bibitem[\protect\citeauthoryear{Preston}{1974}]{Preston74} Preston G.W., 1974, A\&AS, 12, 257
%Astron. Astrophys. Suppl. Ser.

%\bibitem[\protect\citeauthoryear{Reed \& Welch}{1988}]{Reed+Welch88} Reed L.G., Welch G.A., 1988, AJ, 95, 151

%\bibitem[\protect\citeauthoryear{Raskin et al.}{2011}]{Raskin+11} G. Raskin G., Van Winckel H., Hensberge H. et al., 2011, A\&A, 526, 69
%Raskin, G.; van Winckel, H.; Hensberge, H.; Jorissen, A.; Lehmann, H.; Waelkens, C.; Avila, G.; de Cuyper, J.-P.; Degroote, P.; Dubosson, R.; Dumortier, L.; Frémat, Y.; Laux, U.; Michaud, B.; Morren, J.; Perez Padilla, J.; Pessemier, W.; Prins, S.; Smolders, K.; van Eck, S.; Winkler, J.

%\bibitem[\protect\citeauthoryear{Renson \& Manfroid}{2009}]{Renson+Manfroid09} Renson P., Manfroid J., 2009, A\&A, 498, 961

\bibitem[\protect\citeauthoryear{Renson \& Catalano}{2001}]{Renson+Catalano+01} Renson P., Catalano F.A., 2001, A\&A, 378, 113

%\bibitem[\protect\citeauthoryear{Renson et al.}{1991}]{Renson+91} Renson P., Gerbaldi M., Catalano F. A., 1991, A\&AS, 89, 429

\bibitem[\protect\citeauthoryear{Ricker et al.}{2015}]{Ricker+15} Ricker G.R., Winn J.N., Vanderspek R., et al., 2015, Journal of Astronomical Telescopes, Instruments and Systems, 1, 014003

%You can refer to the full collection of TESS-DATA-ALERTS data products using the DOI http://dx.doi.org/10.17909/t9-wx1n-aw08.
\bibitem[\protect\citeauthoryear{Ricker \& Vanderspek}{2018}]{Ricker+Vanderspek} Ricker G., Vanderspek R., 2018, doi:10.17909/t9-wx1n-aw08
%Title: Data Products From TESS Data Alerts

%\bibitem[\protect\citeauthoryear{Romanyuk et al.}{2014}]{Romanyuk+14} Romanyuk I.I., Semenko E.A., Kudryavtsev D.O., 2014, Astrophysical Bulletin, 69, 4, 427

%\bibitem[\protect\citeauthoryear{Royer et al.}{2002}]{Royer+02} Royer F., Grenier S., Baylac M.-O., Gomez A.E., Zorec J., 2002, A\&A, 393, 897

%\bibitem[\protect\citeauthoryear{Ryabchikova et al.}{2003}]{Ryabchikova+03} Ryabchikova T., Wade G.A.; LeBlanc F., 2003, in Piskunov N., Weiss W.W., \& Gray D.F., eds, Proc. of IAU Symp. 210, Modelling of Stellar Atmospheres. Astronomical Society of the Pacific, p. 301

%\bibitem[\protect\citeauthoryear{Ryabchikova et al.}{2004}]{Ryabchikova+04} Ryabchikova T., Leone F., Kochukhov O., Bagnulo S., 2004, in Zverko J., \v Zi\v z\v novsk\'y J., \& Adelman S.J., Weiss W.W., eds, Proc. of IAU Symp. 224, The A-Star Puzzle. Cambridge univ. press, Cambridge, p. 580

%\bibitem[\protect\citeauthoryear{Ryabchikova et al.}{2008}]{Ryab+08} Ryabchikova T., Kochukhov O., Bagnulo S., 2008, A\&A, 480, 811

%\bibitem[\protect\citeauthoryear{Ryabchikova et al.}{2015}]{Ryab+15} Ryabchikova T., Piskunov N., Kurucz R.L., Stempels H.C., Heiter U., Pakhomov Yu., Barklem P.S., 2015, Physics Scripta, 90, 5, id. 054005

\bibitem[\protect\citeauthoryear{Saio}{2005}]{Saio+05}	Saio H. 2005, MNRAS, 360, 1022

%\bibitem[\protect\citeauthoryear{Schlafly \& Finkbeiner}{2011}]{Schlafly+Finkbeiner11} Schlafly E.F., Finkbeiner D.P., 2011, ApJ, 737, 103

%\bibitem[\protect\citeauthoryear{Schlafly et al.}{2014}]{Schlafly+14} Schlafly E.F., Green G., Finkbeiner D.P. et al., 2014, ApJ, 789, 15

%\bibitem[\protect\citeauthoryear{Scott et al.}{2015a}]{Scott+15a} Scott P., Grevesse N., Asplund M. et al., 2015a, A\&A, 573, 25

%\bibitem[\protect\citeauthoryear{Scott et al.}{2015b}]{Scott+15b} Scott P., Asplund M., Grevesse N., Bergemann M., Sauval A.J., 2015b, A\&A, 573, 26

\bibitem[\protect\citeauthoryear{Seabold \& Perktold}{2010}]{Seabold+Perktold10} Seabold S., Perktold J., 2010, in van der Walt S. \& Millman J., eds, Proc. of the 9th Python in Science Conference, p. 62

%\bibitem[\protect\citeauthoryear{Shorlin et al.}{2002}]{Shorlin+02} Shorlin S.L.S., Wade G.A., Donati J.-F., et al., 2002, A\&A, 392, 637

%\bibitem[\protect\citeauthoryear{Shulyak et al.}{2004}]{Shulyak+04} Shulyak D., Tsymbal V., Ryabchikova T., St\"{u}tz Ch., Weiss W.W., 2004, A\&A, 428, 993

%\bibitem[\protect\citeauthoryear{Shulyak et al.}{2009}]{Shulyak+09} Shulyak D., Ryabchikova T., Mashonkina L., Kochukhov O., 2009, A\&A, 499, 879


\bibitem[\protect\citeauthoryear{Shultz et al.}{2018}]{Shultz+18} Shultz M.E., Wade G.A., Rivinius Th., Neiner C., Alecian E., Bohlender D., Monin D., Sikora J. 2018, MNRAS, 475, 5144
% B stars
% Shultz, M. E.; Wade, G. A.; Rivinius, Th; Neiner, C.; Alecian, E.; Bohlender, D.; Monin, D.; Sikora, J.; MiMeS Collaboration; BinaMIcS Collaboration

\bibitem[\protect\citeauthoryear{Silvester et al.}{2014}]{Silvester+14} Silvester J., Kochukhov O., Wade G.A. 2014, MNRAS, 440, 182

\bibitem[\protect\citeauthoryear{Sikora et al.}{2019}]{Sikora+19} Sikora J., David-Uraz A., Chowdhury S., Bowman D.M., Wade G.A., Khalack V., Kobzar O., Kochukhov O., et al., 2019, MNRAS, 487, 4695

\bibitem[\protect\citeauthoryear{Smalley et al.}{2017}]{Smalley+17} Smalley B., Antoci V., Holdsworth D.L., Kurtz D.W., Murphy S.J., De Cat P., Anderson D.R., Catanzaro G., et al. 2017, MNRAS, 465, 2662

%\bibitem[\protect\citeauthoryear{Smith}{1996}]{Smith1996} Smith K.C., 1996, Ap\&SS, 237, 77
%Asptrophys. \& Space Sci.

%\bibitem[\protect\citeauthoryear{Smith \& Dworetsky}{1993}] {Smith+Dworetsky93} Smith K.C., Dworetsky M.M., 1993, A\&A ,274, 335

%\bibitem[\protect\citeauthoryear{Sokolov}{1998}]{Sokolov+98} Sokolov N.A., 1998, A\&AS, 130, 215

\bibitem[\protect\citeauthoryear{Stassun et al.}{2018}]{Stassun+18} Stassun K.G., Oelkers R.J., Pepper J., Paegert M., De Lee N., Torres G., Latham D.W., Charpinet S., et al., 2018, AJ, 156, 102

\bibitem[\protect\citeauthoryear{Stassun et al.}{2019}]{Stassun+19} Stassun K.G., Oelkers R.J., Paegert M., Torres G., Pepper J., De Lee N., Collins K., Latham D.W., et al., 2019, AAS, submitted

\bibitem[\protect\citeauthoryear{Stibbs}{1950}]{Stibbs+50} Stibbs D.W.N. 1950, MNRAS, 110, 395

\bibitem[\protect\citeauthoryear{Stift \& Alecian}{2012}]{Stift+Alecian12} Stift M.J., Alecian G., 2012, MNRAS, 425, 2715

\bibitem[\protect\citeauthoryear{Stift \& Alecian}{2016}]{Stift+Alecian16} Stift M.J., Alecian G., 2016, MNRAS, 457, 74

%\bibitem[\protect\citeauthoryear{Str\"{o}mgren}{1966}]{Stromgren66} Str\"{o}mgren B., 1966, ARA\&A, 4, 433

%\bibitem[\protect\citeauthoryear{Thiam et al.}{2010}]{Thiam+10} Thiam M., LeBlanc F., Khalack V., Wade G.A., 2010, MNRAS, 405, 1384

\bibitem[\protect\citeauthoryear{Townsend \& Teitler}{2013}]{GYRE} Townsend R.H.D., Teitler S.A., 2013, MNRAS, 435, 3406

%\bibitem[\protect\citeauthoryear{Wesselink et al.}{1972}]{Wesselink+72} Wesselink A.J., Paranya K., De Vorkin K., 1972, A\&AS, 7, 257

%\bibitem[\protect\citeauthoryear{Wright \& Eastman}{2014}]{Wright+Eastman+14} Wright J.T., Eastman J.D., 2014, PASP, 126, 838

%\bibitem[\protect\citeauthoryear{Zabriskie}{1977}]{Zabriskie77} Zabriskie F.R., 1977, PASP, 89, 561

%\bibitem[\protect\citeauthoryear{\v{Z}i\v{z}\v{n}ovsk\'{y}}{1980}]{Ziga80} \v{Z}i\v{z}\v{n}ovsk\'{y} J., 1980, BAICz, 31, 300

%\bibitem[\protect\citeauthoryear{\v{Z}i\v{z}\v{n}ovsk\'{y}}{1983}]{Ziga83} \v{Z}i\v{z}\v{n}ovsk\'{y} J., 1983, IBVS, 2366, 1

%\bibitem[\protect\citeauthoryear{\v{Z}i\v{z}\v{n}ovsk\'{y} \& Romanyuk}{1990}]{Ziga+Romanyuk90} \v{Z}i\v{z}\v{n}ovsk\'{y} J., Romanyuk J., 1990, BAICz, 41, 118

%\bibitem[\protect\citeauthoryear{Y\"{u}ce et al.}{2011}]{Yuce+11} Y\"{u}ce K., Adelman S.J., Gulliver A.F., Hill G., 2011, AN, 332, No.7, 681

\end{thebibliography}

%\section{Results}
%\label{analysis}

\section{Discussion}
\label{discus}

%Talk about the precision of period determination. Compare this period with the one from the Hipparcos-2 catalogue!!!

The detected photometric variability of HD~27463 with a period of $P$ =~2.834274 $\pm$ 0.000008 d (see Table~\ref{tab2_All} and Fig.~\ref{fig4}) can be explained in terms of stellar rotation with co-rotating surface abundance/brightness patches, likely due to the presence of a surface magnetic field.
%within the oblique magnetic rotator model \citep{Stibbs+50}.
In a hydrodynamically stable stellar atmosphere even a relatively weak magnetic field can amplify the atomic diffusion and lead to horizontal and vertical stratification of chemical abundances \citep{Alecian+Stift10,Stift+Alecian12,Stift+Alecian16}. Usually the magnetic CP stars host magnetic fields that are dominated by  a dipolar component with strengths on the order of several kG or less, which remains stable over long timescales \citep{Silvester+14,Kochukhov+19}. The magnetically sensitive line profiles
Fe\,{\sc ii} 6147\AA, 6149\AA\, visible in the FEROS spectra of HD~27463 are not visually split due to the Zeeman effect.
Taking into account the derived value of v$\sin{i} =  27 \pm 2$ km~s$^{-1}$, it seems that the magnetic field modulus might be of order of a few kG in the stellar atmosphere of this star. It is significant enough to cause the formation of abundance patches in the stellar atmosphere of HD~27463 taking into account its spectral classification as an Ap EuCr(Sr) star  \citep{Houk+Cowley+75}. A preliminary analysis of the two FEROS spectra shows that the profiles of Si and Ca lines are strongly variable, which confirms the Ap nature of this star.
%Use other reference because \citep{Shultz+18} talk about the hot Bp stars.
The patches of enhanced abundance of different metals usually are located close to the magnetic poles or along the magnetic equator \citep{Luftinger+10,Kochukhov+19,Mathys+19} and contribute to the variability of spectral line profiles and the photometric flux in phase with the stellar rotation. Assuming the oblique magnetic rotator model \citep{Stibbs+50} in the case of HD~27463 the overabundance patches %located at the magnetic poles
will generate a signal at the first harmonic of the rotational frequency. The derived pattern of stellar variability in the low frequency domain is in good accordance with the predictions of the oblique magnetic rotator model (see Fig.~\ref{fig4}).
%and HD~27463 may possess a considerable %relatively strong magnetic field

%Describe determination of the mean surface magnetic field <Bs> from the analysis of Fe\,{\sc II} 6147\AA, 6149\AA\, lines \citep{Khalack+Landstreet12}.

The lower panel of Fig.~\ref{fig4} indicates that HD~27463 shows high-frequency pulsations (40 < $\nu$ < 70~d$^{-1}$)
%at the spectral region from 40 to 70~d$^{-1}$
with amplitudes less than 0.5 mmag. Such high-frequency pulsation modes are typically found in hotter $\delta$~Scuti stars \citep{Smalley+17,Bowman+Kurtz18a}. This is because the heat-engine driving mechanism operating in the He\,{\sc ii} ionisation zone is closer to the surface and is more efficient at exciting higher radial orders, hence higher frequencies, in hotter $\delta$~Scuti stars \citep{Pamyatnykh00,Dupret+04}.
\citet{Cunha+19} and \citet{Sikora+19} have also classified this star as a suspected $\delta$~Scuti variable. Therefore, our spectroscopic parameters and subsequent modelling results are in agreement with this theoretically predicted and observed relationship.

Many $\delta$ Scuti stars are known to show a long-term variability of their pulsation amplitudes and frequencies \citep{Breger00,Bowman16}.
%In this study we have carried out an analysis the amplitude modulation for the detected frequencies of stellar pulsation.
To study possible amplitude variability in HD~27463, we employed the methodology of \citet{Bowman+16}, such that the amplitude and phase of a pulsation mode was tracked at fixed frequency using linear least-squares over the duration of the \textit{TESS} light curve. To achieve this, the %\textit{TESS}
light curve of HD~27463 was divided into bins of 50~d with a step of 1~d, and amplitude and phase of a pulsation mode were optimized using linear least-squares in each bin for the frequency determined using the entire light curve. In our analysis, the zero-point of the time used in the calculation of phases was chosen as the approximate midpoint of the entire \textit{TESS} light curve (i.e. BJD 2471340.0). The 1-$\sigma$ uncertainties for the amplitude and phase values were obtained from the least-squares fit.
%HD~27463 shows many pulsation modes at the frequencies that are closely-spaced (see lower panel at Fig.~\ref{fig4}). A major obstacle in studying amplitude modulation in $\delta$ Scuti pulsations, is that such pulsation modes cover a large frequency range and simultaneously often have closely-spaced frequencies. The combined effect of these modes creates strong beating patterns that can dominate a tracking plot if only a short moving window is used. HD~27463 has been observed by the \textit{TESS} in five consecutive sectors that cover almost 5 months and we can use a window of 50 days and a moving step of 1 day to resolve beating patterns.

\begin{figure}
%\begin{center}
%\vspace{-110px}
%\hspace{60px}
%\includegraphics[width=3.2in]{F59.7AmplitudeVar.eps}
\includegraphics[width=3.2in]{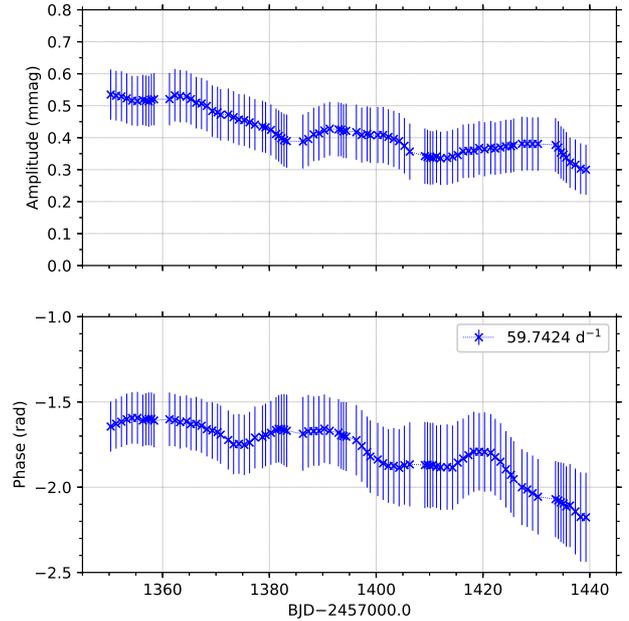}
%\vspace{110px}
\caption{Variability of the amplitude and phase with time for stellar pulsation detected at the frequency $\nu = 59.7424$~d$^{-1}$. 1-$\sigma$ uncertainties are determined from the least-squares fit.}
%d$^{-1}$.}
%\end{center}
\label{fig5}
\end{figure}

Employing this technique we were able to detect marginal amplitude and phase modulation given the relatively large uncertainties for the pulsation modes in HD~27463. An example of the amplitude and phase modulation for the pulsation mode frequency at 59.2474~d$^{-1}$ is shown in Fig.~\ref{fig5}. Other pulsation modes with a relatively high amplitude (see Table~\ref{tab2_All}) show similar behaviour. Our detection of amplitude and phase modulation for pulsation modes with relatively high amplitudes is in agreement with the findings of \citet{Bowman+16} that amplitude modulation is common in $\delta$~Scuti stars with long-term amplitude modulation occurring on time scales much longer than the rotation period.
%Such long-term amplitude modulation is not observed in roAp stars (see \citealt{Cunha+19}).
Some pulsation modes in HD~27463 have very small amplitudes. In this case, it is difficult to search for an amplitude modulation taking into account that the error-bars derived from a non-linear least squares fit are of order 0.01 mmag.

%HD 27463, but the \textit{TESS} light curve currently does not allow us to disprove a null hypothesis that there is no amplitude modulation. For reference, I used a window of 50 days and a moving step of 1 day to resolve beating patterns (see below).

%Therefore, from my preliminary analysis, I am not confident that the amplitude modulation is a significant enough to warrant a discussion in a paper on this star.

We have also analysed the 26 frequencies of detected stellar pulsations located in the frequency range between 40 and 70~d$^{-1}$ (see Table~\ref{tab2_All}) that result in best fitting large  frequency separation around 3.3~d$^{-1}$ which is used as a constraint to model %in modelling of
pulsation modes (see Fig.~\ref{fig:bestfit}). The grid of stellar evolution models has been calculated with {\sc mesa} version 11554 \citep{Paxton+11, Paxton+13, Paxton+15, Paxton+18, Paxton+19} including the effects of convective overshoot and metallicity, while the simulation of linear adiabatic pulsation frequencies for each main sequence model has been carried out using the code {\sc gyre} \citep{GYRE}. Our best fitting model corresponds to an overshoot parameter %the overshoot
of $f_{ov} = 0.014$ and results in values of global stellar parameters that are very close to those reported in the TIC and to those derived in this paper from the simulation of Balmer line profiles (see Table~\ref{tab1}).
This model was determined to have an age of 5.0 $\pm$ 0.4 $\times 10^8$~yr, which corresponds to a core hydrogen fraction of 0.33.

We have found that the large frequency separation is not a sensitive function of the convective core overshoot, as shown in Fig.~\ref{fig:DnuDistribution}. The spectroscopic constraints on fundamental parameters, especially $T_{\rm eff}$ and metallicity, in addition to the observed large frequency separation serves to narrow down the number of models, which must then be compared to the observed echelle ridges (see Figs.~\ref{fig:bestfit},~\ref{fig:splitting}) and the individual frequencies (see Fig.~\ref{fig:individual}) to determine the best fitting parameters.

The radial orders inferred from this work are high for $\delta$~Scuti stars, which are typically observed to have low ($1 \leq n \leq 7$) radial orders. It has been shown by \citet{Antoci+14} that moderate ($7 \leq n \leq 15$) radial orders can be excited in $\delta$~Scuti stars by means of turbulent pressure, although this pulsation excitation mechanism is most efficient in the centre of the classical instability strip. HD~27463 has been shown in this work to be closer the predicted blue edge of the classical instability strip instead of the centre, with an effective temperature of approximately 8700~K. Nevertheless, we conclude that turbulent pressure must play a role, at least in part, in the excitation of such high frequency pulsation modes in HD~27463.

%Binary discussion
Taking into account that HD~27463 is a long period ($\sim$370 yr) visual binary  with a separation of $\sim$0.3 arcsec \citep{Baize+Petit89,Hartkopf+12} we cannot exclude that the detected pulsations are produced by the weaker secondary companion. Having its visual magnitude only 0.43 less than the one of the Ap primary, this companion can also fall in the $\delta$~Scuti instability region and contribute to the observed by \textit{TESS} light curve of HD~27463. Comparison of the two FEROS spectra does not show clear evidences of contribution from the secondary companion. Nevertheless, to estimate parameters of this binary and to clarify the nature of the secondary companion we require acquisition of additional high-resolution and high signal-to-noise spectra of HD~27463.

%For chemically-peculiar stars, the presence of a strong large-scale magnetic field, with a polar magnetic field strength of $B_p$>~1\,kG, is expected to suppress the $\kappa$-mechanism and the excitation of low-radial order pressure modes \citep{Saio+05}.

We have confirmed here that HD~27463 is an Ap star. Excitation of high-radial order magneto-acoustic modes is expected and are observed in some Ap stars (the roAp stars), %observed in roAp stars,
which possess a global magnetic field usually of dipolar structure \citep{Kurtz82,Kurtz+06,Bigot+Kurtz11}. Therefore,
higher resolution spectroscopic and/or spectropolarimetric observations are also required to detect the presence of a strong large-scale magnetic field in HD~27463 to say conclusively whether the star is {\textit{just}} a chemically peculiar $\delta$~Scuti star, or whether it can be called a roAp star.
Our estimation of the upper limit on the magnetic field modulus is not in contradiction with the results of \citet{Saio+05}, who's models require polar magnetic field strengths of $\sim0.8$~kG or more to suppress $\delta$~Scuti pulsations. However,
a precise measurement of the magnetic field in HD~27463 is required to determine the exact driving mechanism in this star and the influence of the magnetic field on the pulsation modes.

The distinction between roAp stars and chemically peculiar $\delta$~Scuti stars based on photometric data alone is becoming less clear (e.g. \citealt{Balona+19}). Typically, an Ap star showing rapid pulsations was historically sufficient to classify
%were sufficient credentials to be classed as
a roAp star. However, the roAp stars can show rotationally split sidelobes in the amplitude spectra of their light curves which are a result of oblique pulsation (e.g. \citealt{Kurtz82,Bigot+Kurtz11}). These sidelobes arise as the pulsation axis in the roAp stars is closely aligned with the magnetic axis which is inclined to the rotational one. To detect the rotational sidelobes, the line of sight to the pulsation needs to be favourable, and the rotation period shorter than the observation period. There is no indication of rotationally split sidelobes in HD~27463 which could mean we are unfortunate in the geometry or the pulsations are not oblique i.e. the pulsation and rotation axes are aligned as in chemically normal $\delta$~Scuti stars.

To conclusively say whether HD~27463 is a chemically peculiar $\delta$~Scuti star or a roAp star and to clarify the nature of the secondary companion, extensive high-resolution spectroscopic and/or spectropolarimetric observations are required. %to detect the presence of a strong large-scale magnetic field.
These data, along with further detailed modelling of star which takes into account the magnetic field, are needed to distinguish between these two classes of pulsator, and to investigate to what extent the magnetic field plays a role in this star and others with similar characteristics. Nonetheless, the rich frequency spectrum of HD~27463 including radial and non-radial modes spanning several radial orders demonstrates the exciting prospects for asteroseismic studies of intermediate-mass stars with \textit{TESS}.

%For chemically-peculiar stars, the presence of a strong large-scale magnetic field is expected to suppress the {\bf k-mechanism} and the excitation of low-radial order pressure modes \citep{Saio+05}. Instead, the excitation of high-radial order magneto-acoustic modes are expected and observed in roAp stars. The detection of rotationally-split multiplets of non-radial pulsation modes with high radial orders in a (magnetic) chemically-peculiar star is the discriminating factor between roAp and $\delta$~Scuti stars. Therefore the determination of the exact driving mechanism in HD~27463 requires the (non-)detection of a strong large-scale magnetic field in HD~27463, but this is beyond the scope of the present work. Nonetheless, the rich frequency spectrum of HD~27463 including radial and non-radial modes spanning several radial orders demonstrates the exciting prospects for asteroseismic studies of intermediate-mass stars with TESS.

%From our grid of models calculated for different stellar masses and overshoot parameters we have found that the difference in the large separation produced our models with respect to the observed value does not depend much on the overshoot (see Fig.~\ref{fig:DnuDistribution}). Therefore to reach the best fit, we narrowed down the models taking into account the large frequency separation and then tried to fit frequencies at the echelle diagrams (see Figs.~\ref{fig:bestfit} and~\ref{fig:splitting}).

\section*{Acknowledgments}

The authors are thankful to Profs. Don Kurtz, Timothy Bedding, Daniel Huber and the anonymous referee for useful suggestions and comments that helped to improve significantly this work.
V.K., C.L., A.D.U. and G.A.W. acknowledge support from the Natural Sciences and Engineering Research Council of Canada (NSERC).
O.Kobzar and V.K. are thankful to the Facult\'{e} des \'{E}tudes Sup\'{e}rieures et de la Recherch and to the Facult\'{e} des Sciences de l'Universit\'{e} de Moncton for the financial support of this research. O.K. acknowledges support by the Swedish Research Council (project 621-2014-5720) and the Swedish National Space Board (projects 185/14, 137/17). D.L.H. acknowledges financial support from the Science and Technology Facilities Council (STFC) via grant ST/M000877/1.
This paper includes data collected by the \textit{TESS} mission. Funding for the \textit{TESS} mission is provided by the NASA Explorer Program. Funding for the TESS Asteroseismic Science Operations Centre is provided by the Danish National Research Foundation (Grant agreement no.: DNRF106), ESA PRODEX (PEA 4000119301) and Stellar Astrophysics Centre (SAC) at Aarhus University. We thank the \textit{TESS} and TASC/TASOC teams for their support of the present work. This research has made use of the SIMBAD database, operated at CDS, Strasbourg, France. Some of the data presented in this paper were obtained from the Mikulski Archive for Space Telescopes (MAST). STScI is operated by the Association of Universities for Research in Astronomy, Inc., under NASA contract NAS5-2655. The research leading to these results has received funding from the European Research Council (ERC) under the European Union's Horizon 2020 research and innovation programme (grant agreement No. 670519: MAMSIE).
%Some of the data presented in this paper were obtained from the Mikulski Archive for Space Telescopes (MAST). STScI is operated by the Association of Universities for Research in Astronomy, Inc., under NASA contract NAS5-26555. We are using here the TESS-DATA-ALERTS data products obtained for the Sectors 1-5.
This study is also partially based on data obtained from the ESO Science Archive Facility under request number Khalack399537.

\appendix
\section{Degeneracy of the large frequency separation}
\label{append}

%When I refit the results, I made my initial guess exactly 6.6, and then adjusted the frequency spacing to see if it improved the echelle diagram, and got something that looked good at 6.64.  I gather Tim did not give us an actual value, and has not been very forthcoming when asked.  If we do get an actual value from Tim, I can easily re-run the fit with that value, but I don't think it's going to change the actual results much.
%Figure 6 should remain unchanged.  Perhaps we could just add a statement that Bedding et al (2019) find that the large separation for this star is larger.  We have attempted to fit HD27463 assuming a large separation near 6.6, and the results are discussed in Appendix A. Then we can include the figure I sent yesterday and a bit of discussion in the appendix.

\begin{figure}
%\vspace{-110px}
%\hspace{40px}
\includegraphics[width=3.5in]{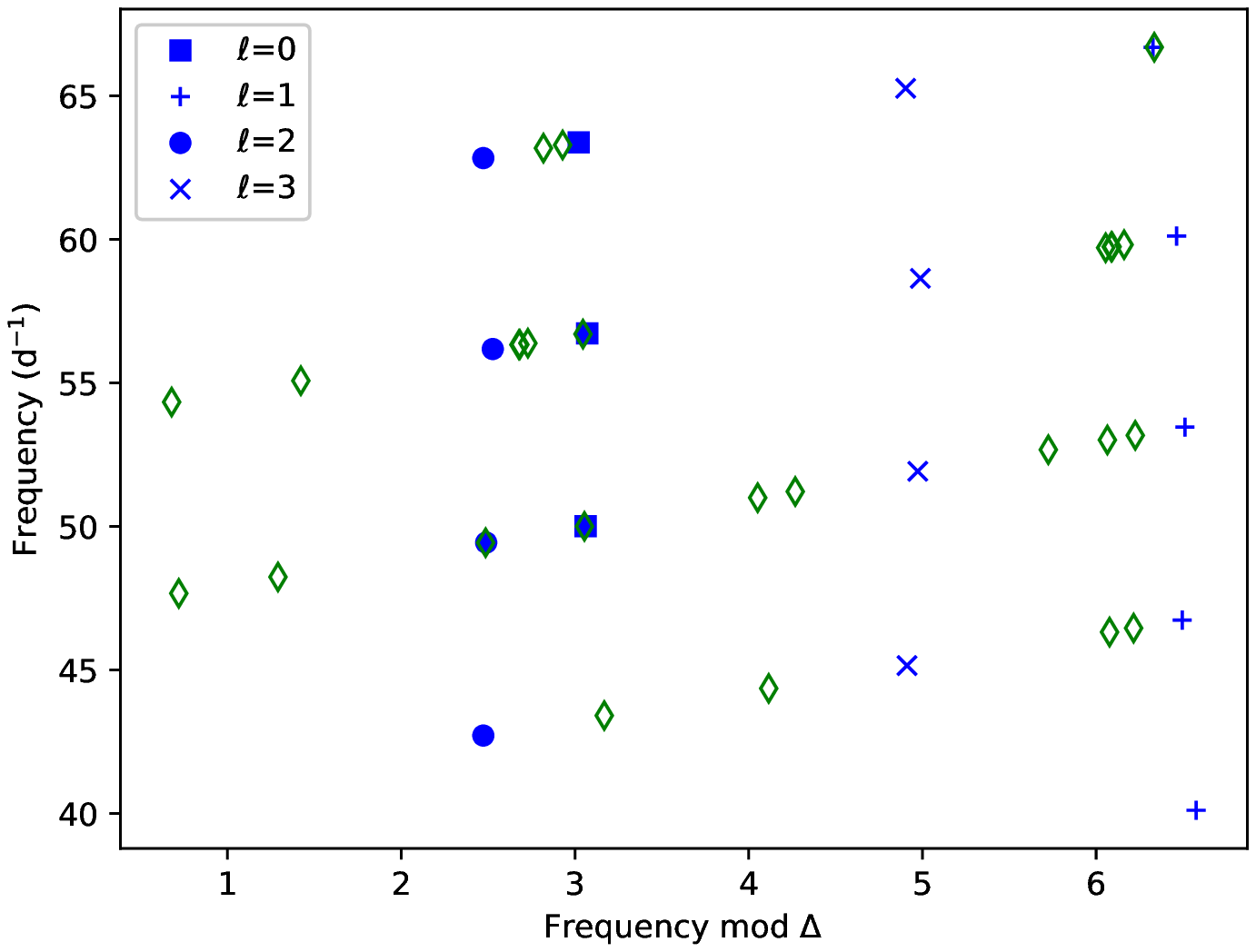}
%\vspace{110px}
\caption{\label{fig:largeA} The echelle diagram for the best fitting model and the observed frequencies, both plotted with a large frequency separation of $\Delta \nu_0$= 6.707~d$^{-1}$.  The blue points show the model frequencies for $\ell = 0$ (squares), $\ell = 1$ ( pluses), $\ell = 2$ (circles), and $\ell$ = 3 (crosses), while the green open diamonds specify the data derived from observations. }
%The  model frequencies plotted here correspond to radial orders of $n = 11 - 19$.}
\end{figure}

\begin{figure}
%\vspace{-110px}
%\hspace{40px}
\includegraphics[width=3.5in]{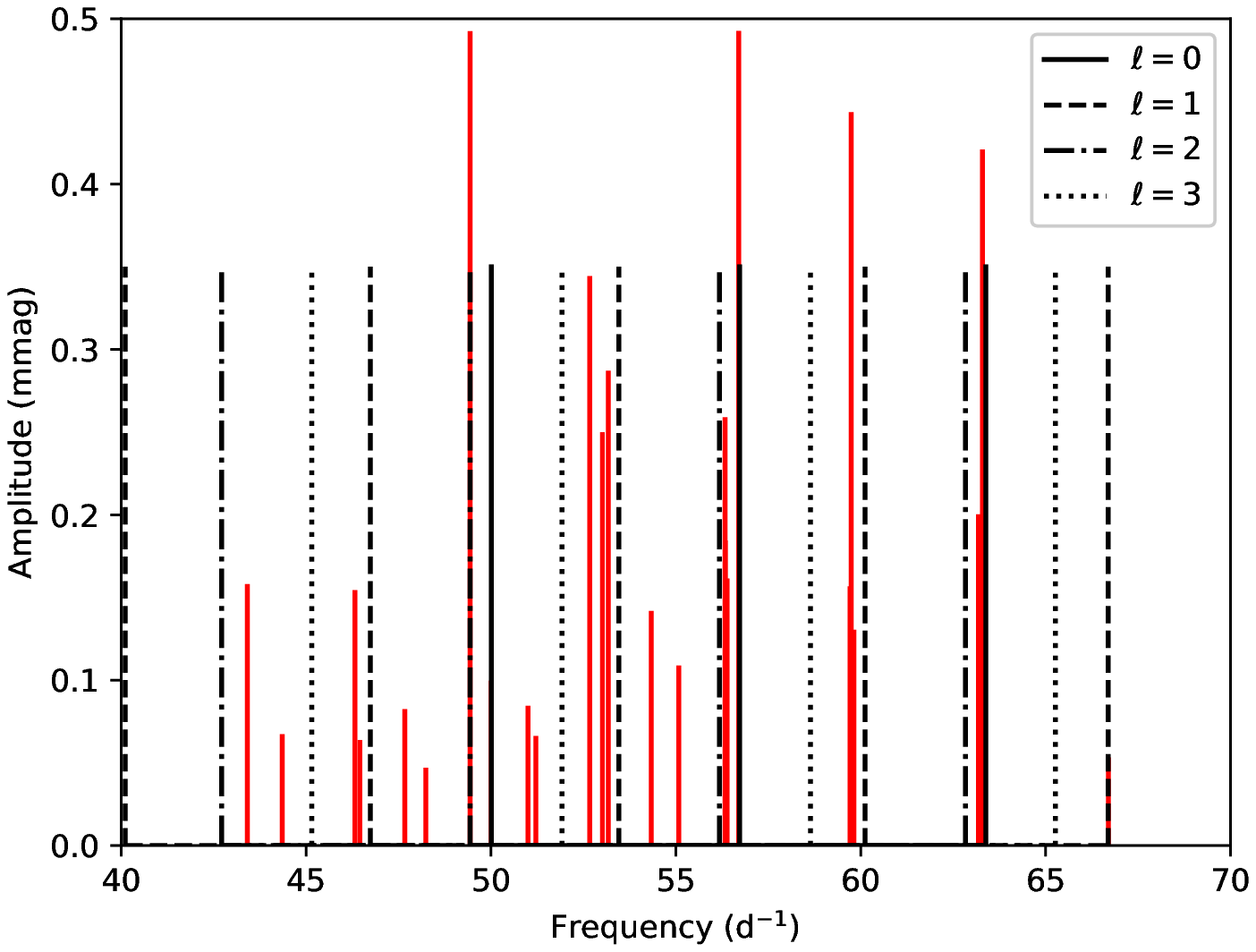}
%\vspace{110px}
\caption{\label{fig:individual6} Individual frequency matches for the best fitting model using the large frequency separation of $\Delta \nu_0= 6.707$~d$^{-1}$. The red lines are the observed frequencies plotted versus amplitude, while the black lines are the $\ell = 0$ (solid), $\ell=1$ (dashed), $\ell=2$ (dot-dashed), and $\ell=3$ (dotted) model frequencies.}
\end{figure}

\label{lastpage}

\end{document}